\documentclass[a4paper, 12 pt, reqno]{article}

\usepackage{amsmath,amssymb, amsthm}
\usepackage{mathrsfs,enumerate,latexsym,graphicx}
\usepackage[hidelinks]{hyperref}
\usepackage{fullpage}
\usepackage{abstract}
\usepackage{setspace}
\usepackage{mathtools}
\usepackage{float}
\usepackage{multirow} 
\usepackage{makecell}
\usepackage{booktabs}
\usepackage{tabularx}
\usepackage{bbm}
\usepackage{doi}
\usepackage{enumitem}
\usepackage{natbib}
\usepackage{calrsfs}
\usepackage{verbatim}
\usepackage[margin=1in]{geometry}
\usepackage{setspace}
\usepackage{array}
\usepackage{footmisc}
\newcolumntype{H}{>{\setbox0=\hbox\bgroup}c<{\egroup}@{}}

\expandafter\def\expandafter\normalsize\expandafter{%
	\normalsize
	\setlength\abovedisplayskip{6pt}
	\setlength\belowdisplayskip{6pt}
	\setlength\abovedisplayshortskip{4pt}
	\setlength\belowdisplayshortskip{4pt}
}

\renewcommand{\vec}[1]{\mathbf{#1}}
\renewcommand{\vec}[1]{\boldsymbol{#1}}
\DeclareMathOperator*{\E}{\mathbb{E}}

\usepackage[usenames,dvipsnames,svgnames,table]{xcolor}

\DeclarePairedDelimiterX\Basics[1](){ #1}

\newcommand{\Gmat}{\mathbf{G}}

\newcommand{\suppliers}{N^{-}}
\newcommand{\customers}{N^{+}}
\newcommand{\indeg}{d^{-}}
\newcommand{\outdeg}{d^{+}}
\newcommand{\cond}{\mathbf{z}}

\usepackage{xurl}
\title{Learning to Import through Production Networks \vspace{-20pt}}
	\vspace{-80pt}
 \author{Kenan Huremovi\'{c} \quad   Federico Nutarelli \quad Francesco Serti\quad \\ Fernando Vega-Redondo\footnote{
 \begin{doublespace}
     K. Huremovi\'{c}, IMT School for Advanced Studies Lucca, kenan.huremovic@imtlucca.it; F. Nutarelli, IMT School for Advanced Studies Lucca, federico.nutarelli@imtlucca.it; F. Serti, IMT School for Advanced Studies Lucca, francesco.serti@imtlucca.it; F. Vega-Redondo, Chinese University of Hong Kong, f.vega.redondo@gmail.com. %We are very grateful to Asier Mariscal and Adam Szeidl for their helpful comments and suggestions, as well as to numerous seminar and conference participants (NETEF 2018, INTECO 2018, 24th CTN 2019, ISGEP 2022,  Universidad de Navarra, Universidad de Deusto, Universitat de les Illes Balears). The usual disclaimer applies.
 \end{doublespace}}} 
\begin{document}

\begin{titlepage}

\clearpage\maketitle
\thispagestyle{empty}

\vspace{-40pt}
\begin{abstract}
%\begin{singlespace}	
		\vspace*{-15pt}

\noindent Using administrative data on the universe of inter-firm transactions in Spain, we show that firms learn to import from their domestic suppliers and customers.
 Our identification strategy exploits the panel structure of the data, the firm-time variation across import origins, and the network structure.
%Our identification strategy provides explicit conditions under which, controlling for time-varying observable and unobservable characteristics of a firm and its partners, the lagged importing status of its second-order contacts, who are not directly connected to the firm, can serve as a valid instrument for the import status of its first-order contacts within a specific geographic area of import origin.
We find evidence of both upstream and downstream network effects, even after accounting for sectoral and spatial spillovers. We estimate that an increase of 10 percentage points in the share of suppliers (customers) that are importing from a given region increases the probability of starting importing from that region by 10.7\% (19.2\%). % For an average firm, an additional supplier (customer) that imports from a given region increases the likelihood of starting to import from that region by 10\% (17\%). %Connections with geographically distant firms provide more useful information to start importing.
Connections with geographically distant domestic firms provide more useful information to start importing. 
Larger firms are more responsive to this information but less likely to disseminate it. %less effective at disseminating it.

% we find that an increase of 10 percentage points in the share of suppliers (customers) that are importing from a given origin increases the probability of starting importing by 10.7\% (19.2\%)

%Old one 
  % \noindent Using data on the Spanish firm-level production network we
  % show that firms learn about international trade opportunities and related business know-how from their production network peers. Our identification strategy leverages the panel structure of the data, import origin variation, and network structure. We find evidence of both upstream and downstream network effects, even after accounting for sectoral and geographical spillovers.  Larger firms are better at absorbing valuable information but worse at disseminating it. Connections with geographically distant firms provide more useful information to start importing.
  %\kenan{Sentence or two about punchline results}
%\end{singlespace}
\end{abstract}
\vspace{-10pt}

\hspace{15pt}\textbf{Key Words:}  Production network, Learning, Spillovers, Import.

\hspace{15pt}\textbf{JEL Codes:} D22, D83, F14, L14.
\end{titlepage}
\newpage
   
    % \begin{table}[h]
    % 	 \begin{center}
    % 	\begin{tabular}{|l|l|}
    % 		\hline
    % 		D22 & Firm Behavior: Empirical Analysis \\
    % 		\hline
    % 		D83 & Search, Learning, Information and Knowledge,\\
    % 		&Communication, Belief, Unawareness\\
    % 		\hline
    % 		F14 & Empirical Studies of Trade\\
    % 		\hline
    % 		L14 & Transactional Relationships, Contracts and Reputation,\\
    % 		& Networks\\
    % 		\hline 
    % 	\end{tabular}
    % \end{center}
    % \end{table}

\section{Introduction} \label{sec:Intro}

A large body of literature emphasizes the importance of firm heterogeneity in terms of productivity and destination-specific entry costs for explaining the selection of firms in international markets. After concentrating on the firm-level export behaviour \citep{melitz2003impact, eaton2011anatomy}, the research agenda has also focused on firms' international sourcing decisions \citep{halpern2015imported, antras2017margins, lu2024firms}. Understanding the determinants of firms' importing behaviour is crucial because intermediate inputs represent approximately two-thirds of world trade \citep{johnson2017portrait}, and imports are important for firm performance (for instance \cite{amiti2007trade, kasahara2008does,  halpern2015imported, bernard2019production}, among many other papers). 

%Firm-specific and country-specific factors are commonly identified as key determinants of importing decisions (@ADD CITATION HERE, 1-2 PAPERS). However, these factors alone fail to fully account for the significant variation observed in importing along the extensive margin (\cite{antras2017margins}). In their structural analysis, \cite{antras2017margins} highlight that a firm’s decision to import is strongly influenced by the presence of existing foreign suppliers from different countries and that accounting for this mechanism greatly improves the model’s ability to replicate empirical patterns.
%while certain country characteristics may make some countries particularly appealing to all firms,
In their structural analysis of importing \cite{antras2017margins} highlight the importance of accounting for heterogeneity at the firm-import origin (i.e., country) level to explain the observed empirical patterns of importing at the extensive margin. 
%to improve the fit of their model is necessary to go beyond common country fixed costs and consider the firm-country dimension, because significant firm-specific idiosyncrasies in sourcing strategies are present.
In this paper, we examine firms' decisions to start importing from a specific origin, focusing on an unexplored source of firm heterogeneity that operates at the firm-origin level: a firm's position in the domestic production network. We hypothesize that when exchanging goods and services, firms may also share valuable information about importing from specific countries or geographic regions. 
 This includes information about potential foreign trading partners, such as price, quality, and trustworthiness, as well as specific know-how related to the informational component of fixed trade costs, such as understanding institutional conditions, corporate culture, business practices, and other non-tariff barriers affecting market access.%\footnote{In the model of \cite{antras2017margins}, our proposed mechanism could be represented by incorporating the firm's position within the domestic network as a determinant of country-specific fixed costs. The decision to start importing depends on both the fixed costs and the expected revenue gains. Empirically determining whether peer information reduces these fixed costs or facilitates access to intermediates that enhance expected gains (e.g., inputs with lower expected quality-adjusted prices) is beyond the scope of this paper and left for future research. A formal theoretical model of the mechanism we empirically examine is available upon request.} 
 % endowed with heterogeneous and incomplete information about trading opportunities and know-how
 
%To test these hypotheses, 
We empirically investigate the diffusion of information about importing through the domestic production network using a dataset provided by the Spanish Tax Agency (AEAT), which contains data gathered from Value Added Tax (VAT) declarations. This dataset includes anonymized information about the basic characteristics of the whole population of Spanish firms, together with the value of their imports for two aggregate geopolitical areas (EU and extra-EU) and all annual domestic transactions between them in an amount larger than 3,005 Euro. By leveraging this dataset, we construct the empirical Spanish domestic firm-level production network for each year during the 2010--2014 period. %The observed transactions between firms can be used to define an interaction matrix that determines a firm's exposure to information coming from its peers that are importing, distinguishing between providers and customers.
We then empirically examine whether the (geographical, i.e. area-specific) import experience of a firm's domestic trade partners, differentiating between its providers and customers, is relevant for explaining its decision to start importing (from this area).\footnote{We focus on firms' importing behaviour at the extensive margin, i.e., import starters, and leave the investigation of the intensive margin for future work.}

We distinguish between spillovers coming from suppliers (\textit{downstream effect}) and customers (\textit{upstream effect}) because they may have different incentives to share such information with the firm, and they may also transmit different information. On the one hand, a firm may be willing to share information on potential import opportunities with its suppliers in order to enhance the quality or decrease the cost of the sourced inputs, while suppliers may be reluctant to disclose such information to their customers, as it may put them at risk of being supplanted by foreign providers.\footnote{This is consistent with the estimates in the literature suggesting that the intermediate inputs tend to be substitutes (see, for instance, \cite{carvalho2021supply,huremovic2023production})} On the other hand, the relevance of information about potential foreign suppliers is likely to be greater when it comes from suppliers rather than customers, given their upstream position in the production chain.\footnote{For instance, \cite{carvalho2014input} and \cite{chaney2018gravity} (in the online appendix) highlight the importance of existing suppliers in finding potentially useful inputs.} Hence, a priori, it is unclear whether one should expect forward or backward linkages to be a relatively more important source of information spillovers related to importing opportunities and know-how. %Therefore, in our analysis, we distinguish between spillovers coming from suppliers (\textit{downstream effect}) and customers (\textit{upstream effect}). 
% In their industry-level empirical analysis, \cite{carvalho2014input} use U.S. input-output tables at the 4-digit level to define the distance between industries and show that the backward distance (i.e., via suppliers) and not the forward distance (i.e., via customers), predicts input adoption  
%General description of what we do

We estimate these peer effects in a linear-in-means framework (i.e. \cite{bramoulle2009identification}), therefore assuming that a firm's decision to start importing from a given origin/area is affected by the firm's characteristics, a weighted average of its peer characteristics, and the weighted averages of the dichotomous importing status (importer or not) of its peers. In our benchmark specification, each supplier (customer) of the firm is weighted equally, and we normalize the weights to add up to 1. Therefore, the weighted average of the importing status of suppliers (customers) is the share of suppliers (customers) that import from a given origin. Unlike the standard linear-in-means setting, we assume that peer effects operate with time lag -- it takes time for a firm to utilize the import-relevant knowledge acquired from its peers. The same assumption is made in the papers by \cite{bisztray2018learning} and \cite{haller2023importer}, which %studies location-based peer effects in importing and 
we discuss later at some length. % and \cite{dhyne2023export}.

%Issues in estimation of linear-in-means
There are several well-known challenges in estimating the linear-in-means model. The most important problem derives from the possible existence of correlated effects. Correlation in outcomes among peers may arise due to endogenous choice of peers or to common shocks. This problem generally occurs when a correlation exists between peers' unobserved characteristics. The second issue is the reflection problem, which prevents separate identification of the impact of peers' outcomes (endogenous peer effects) and peers' characteristics (contextual peer effects) whenever the peer effects are contemporaneous. %\footnote{The reflection problem, however, can generally be addressed in the case of network interactions. See \cite{bramoulle2020peer} for an excellent review of the identification of peer effects in networks.} 
As explained below, we tackle these issues by combining different strategies pursued by the literature addressing identification problems in estimating network peer effects (see \cite{bramoulle2020peer} for an excellent review of this literature). 

%Identification exposition could be improved 

As mentioned, we assume a delay in peer effects since we expect information diffusion to take time. This assumption practically makes the reflection problem inconsequential in our setting since it breaks the simultaneity of endogenous peer effects and a firm's decision to import. 
To deal with the issue of correlated effects, we start from a commonly made assumption that the network is conditionally exogenous, and we apply it to a panel data context, a setting that is surprisingly uncommon in the literature on peer effects in networks%\citep{bramoulle2020peer}
. By exploiting the panel structure of our data, we control for a substantial set of observable and unobservable characteristics.  In our preferred specifications, we control for \textit{firm$\times$year} and \textit{import-origin$\times$firm-industry$\times$firm-location$\times$year} fixed effects. This demanding set of fixed effects accounts for multiple channels through which the issue of correlated effects might influence the estimation of peer effects. It includes geographical and industry-level shocks, as well as potential correlations among unobservable factors impacting peer-importing behaviour that are not origin-specific, such as for instance, demand or productivity factors. 

Our identification of spillovers relies, therefore, on the import-origin variation in the share of importing neighbours that is independent of their geographical and sectoral distribution, and of the time-varying characteristics (but not import-origin specific) of the firm and its neighbours. Consequently, any potential residual threats to identification must manifest at this nuanced level of variation. To address them, we exploit the network structure. In particular, we show how the importing status, at $t-2$,  of a firm's suppliers of suppliers (customers of customers) that are neither that firm's suppliers nor customers can be used as a valid instrument for the firm's suppliers' (customers') importing status at $t-1$. Finally, in order to mitigate the issue of network endogeneity we fix the network by considering only supplier-customer connections that appear in the dataset throughout 2010--2014.

%Finally, in addition to exploiting the panel structure of the data to wash out a wide range of unobservables, we mitigate the issue of network endogeneity by fixing the network. This means that we consider only supplier-customer connections that appear in the dataset throughout 2010--2014. 

In our preferred specification, which combines the most demanding set of fixed effects (i.e., \textit{firm$\times$year} fixed effects and \textit{import-origin$\times$firm-industry$\times$firm-location$\times$year} fixed effects) with our IV strategy,  we find that an increase of 10 percentage points in the share of suppliers (customers) that are importing from a given origin increases the probability of starting importing by 10.7\% (19.2\%). As already explained, \textit{a priori} it is unclear whether one should expect that the downstream diffusion of information going from suppliers to customers should be be greater or smaller than the upstream counterpart flowing from customers to suppliers. 
%a supplier of a firm may be replaced with a foreign provider when the firm starts importing (instead, a customer could indirectly benefit when its supplier uses more productive or cheaper inputs), but the information coming from providers could be more relevant to sourcing new inputs because of their upstream position in the supply chain. 
Indeed, we find that although the point estimate of the upstream effect is nearly twice that of the downstream effect, the difference is not statistically significant. This suggests that the potentially higher relevance of information from suppliers is balanced by their weaker incentives to share the knowledge.  Another interesting finding is that the spillover effect is estimated to be important only for the more specialized import origin-specific knowledge, as we do not find evidence of peer effects for non-origin-specific importing. 

Our analysis reveals significant heterogeneity in upstream and downstream effects. We find evidence that larger and more productive firms absorb and utilize import-relevant information better. They are, strategically or not, less effective in disseminating the information. The spillovers are stronger when coming from firms in the same industry due to the similarity in production technology. Interestingly, connections with domestic peers that are geographically distant provide more useful information for importing. Given the localized nature of production networks, this is consistent with the "strength of weak ties" effect postulated in \cite{granovetter1973strength} in the context of social networks.

The remainder of the paper is organized as follows. Section \ref{sec:Lit} situates our study within the existing literature. Section \ref{sec:data} describes the data, while Section \ref{sec:EmpiricalSpecification} outlines the empirical strategy. Sections \ref{sec:Results} and \ref{ss:Heterogenity} present the results. Finally, Section \ref{sec:Conclusion} summarizes our findings and discusses potential directions for future research. Tables containing the results are provided in the Appendix.

\section{Relation to the Literature} \label{sec:Lit}

%%%%%Spostato 

Since the seminal paper of \cite{rauch1999networks}, trade information frictions and the role of networks in overcoming them have been studied theoretically and empirically \citep{Chaney2016}. Forming a new trade relation typically requires substantial effort in gathering information that is not freely available but is acquired through search and learning efforts. Indeed, to start to trade, firms first need to be aware of the existence of a trading opportunity. Once the potential trading partner has been identified, there are additional obstacles to establishing a successful trade relationship, including learning how to do business in the presence of non-tariff barriers (safety regulations, formal trade procedures, etc.) and issues related to incomplete information \citep{allen2014information}. For example, a firm's decision to start purchasing an input from a new provider is always, to some extent, characterized by uncertainty about the ability of the potential seller to fulfil its needs in terms of price, quality, and delivery \citep{Rauch2003} and about the cost of integrating the outsourced input into the production process.

Knowledge spillovers are an important topic in the international trade literature. The knowledge spillovers associated with importing have been theoretically and empirically explored, although receiving significantly less attention than those associated with exporting.  For instance, 
 % On the theoretical side,
 \cite{allen2014information} shows that, in the context of regional agricultural trade flows in the Philippines, producers/sellers' search costs for acquiring information about market conditions in other locations can explain about half of the observed price dispersion across regions.
 %In related work, \cite{DASGUPTA2018150} build a model in which importers have a limited capability to process information about prices and tend to allocate their scarce resources to countries with lower expected prices. \cite{albornoz2012sequential}, \cite{li2018structural}, \cite{berman2019demand} and \cite{eaton2024search} estimate international trade models in which firms learn about export profitability and the appeal of their products from their own export activities. \cite{monarch2017learning} propose a model in which importers update the information about the quality of their foreign suppliers by observing their accumulated exchanges. More closely 
 Related to our study, a few papers provide theory models in which firms use their connections as a source of information to form new trade relations. For instance, in  \cite{rauch2004network}, agents with networks of foreign contacts have the option to either leverage these networks for their own production or offer them for use by others % In \cite{krautheim2007gravity} firms form information-exchange links with other firms from the same industry, and acquire beneficial information about exporting through these links. 
 %In \cite{chaney2014network}, exporters search for new customers in a given location by using their existing customers from that location in the spirit of \cite{jackson2007meeting}.
 while \cite{chaney2018gravity} provides a stylized model in which symmetric firms can buy information about new suppliers from the current suppliers for a fixed and exogenously given price.

Although importing has been shown to be crucial for determining firm productivity and export outcomes,\footnote{Imported inputs have been recognized as one of the main sources of technological diffusion and, consequently, of productivity growth 
 \citep{coe1995international, keller2002trade}. \cite{Bernard2009} and \cite{castellani2010firms} demonstrate that the heterogeneity among importers is significant, highlighting notable differences between importers and non-importers. The use of foreign intermediate inputs increases productivity due to the higher quality of the foreign technology embodied in imported inputs and to the complementarity between foreign and domestic inputs \citep{amiti2007trade, kasahara2008does, halpern2015imported}. The literature also suggests that the use of foreign intermediates promotes quality upgrading \citep{kugler2009plants, kugler2011prices}, the introduction of new final products \citep{goldberg2010imported, colantone2014new} and R\&D investments \citep{boler2015r}. The literature on firm-to-firm trade has emphasized the importance of having more and better suppliers for the performance of downstream firms both in domestic and international production networks    \citep{chaney2018gravity,bernard2019production,bernard2018networks}.} there are only a few papers that empirically study knowledge spillovers in importing. \cite{lopez2010imports} finds that, for Chilean plants, the probability of importing is increasing in the number of importers in the same region. \cite{bisztray2018learning} use a very fine-grained definition of neighbourhood (i.e., being in the same building) considering all types of imported products in the town of Budapest, while \cite{bekes2020machine} concentrate only on machine imports but analyze the whole territory of Hungary. Both studies provide evidence of import spillovers and highlight that their significance grows with proximity to peers. Additionally, \cite{haller2023importer} shows that removing Multi-Fibre Agreement quotas leads firms to import more in commuting zones with a higher concentration of importing peers. All this literature focuses on spatial spillovers and does not consider information transmission through the production network as a potential determinant of importing. Furthermore, by not studying information propagation via buyer-supplier connections, these papers cannot leverage the intransitivities in the network structure to disentangle peer effects, which we crucially exploit in our identification strategy.

Our paper also relates to the recent literature on firm-to-firm trade.  Focusing on domestic production networks, \cite{bernard2019production} proposes a model in which downstream firms search for suppliers with fixed search costs, while 
%incur fixed costs to observe the exact prices of inputs from upstream firms at specific locations. As these costs decrease, firms expand their search across more locations, increasing the likelihood of finding cheaper suppliers, which in turn enhances firm performance.
in \cite{carvalho2014input}, firms search for new inputs through vertical connections.  
%Similar to \cite{carvalho2014input},  \cite{ bernard2018two}  considers a model in which international firm-to-firm connections are characterized by relationship-specific costs.  %and to the degree of enforceability of international contracts. 
 \cite{eaton2022two} and \cite{eaton2024search} model transnational business relationships by resorting to search frictions and random matching. In these frameworks, search costs depend on search intensity and the number of accumulated successful matches. In \cite{kalina2024}, buyer-supplier matching is characterized by fixed costs paid by the buyer proportional to the number of providers, among which the firm can choose within a given origin country-industry.  
 We contribute to this literature by empirically investigating a specific and important channel through which firms search and find foreign suppliers -- domestic production network.  %si puo aggiungere arkolakis
%In \cite{eaton2022two}, the search efforts of both importers and exporters determine the probability of finding new trade partners, but also market tightness and accumulated success play a role.

%\footnote{Since \cite{bernard2015production} focus on sourcing locations within the domestic production network, average differences in prices for firms located in different locations are related to shipping costs and costs related to the buyer-supplier relationship, but not to technological differences that are more relevant across countries.}

%Finally, in our analysis we use tools developed in the literature on peer effects in social networks (\cite{ bramoulle2020peer}). We also show how the insights obtained from this literature can be used in studying issues in international trade in conjunction with production networks.

%The papers that are closest to ours are \cite{bisztray2018learning} and \cite{dhyne2023export}. In \cite{bisztray2018learning}, the authors also study spillovers in importing across connected firms. Differently from our paper, however, they focus on spatial spillovers using a fine-grained but group-based definition of the neighborhood (i.e., being in the same building) while we study information propagation through buyer-supplier connections. Hence, they cannot exploit the intransitivities in the network structure to identify peer effects, which we crucially use in our identification strategy.

The paper that is closest to ours is \cite{dhyne2023export}. The authors use Belgian data on firm-level production networks to study peer effects in exporting. While both their paper and ours consider information transmission through production networks, our approach differs from theirs in several important ways, a few of which we highlight here. Our focus is on importing -- a significantly less studied topic in international trade. As noted by \cite{antras2017margins}, decisions to import are inherently more complex than decisions to export. In contrast with selling in foreign markets, incorporating foreign inputs into production processes affects marginal costs, with this impact depending intricately on the firm's current input mix. This complexity may partially explain why importing has received less attention in the literature.
%Our focus is on importing -- a significantly less studied topic in international trade. As noted by \cite{antras2017margins}, decisions to import are inherently more complex than decisions to export.
% Incorporating foreign inputs into production processes affects marginal costs, with this impact depending intricately on the firm's current input mix. 
% This complexity may partially explain why importing has received less attention in the literature. 
Furthermore, the incentives for domestic suppliers and customers to share information with other firms depend on whether the information pertains to imports or exports, and so happens with the relevance of that information in each case.\footnote{As discussed before, domestic suppliers may hesitate to disclose potential foreign suppliers to avoid being replaced, while domestic customers might share such information to try to improve the quality or reduce the price of the inputs sourced domestically. Instead, in the case of exports, domestic suppliers may be interested in increasing the sales of their buyers to stimulate their derived demand for the intermediate inputs they sell to them. At the same time, customers may not wish to transmit the knowledge to their suppliers as exporting may increase suppliers' bargaining power in their relationship. 
%Suppose domestic customers reveal information on export opportunities to their suppliers. In that case, the latter will increase their sales and, possibly, their prices (there could be an increase also in their productivity due to learning by exporting effects or simply related increasing returns to scale, but it is unclear to what extent this will translate into lower prices or just higher mark-ups).
Finally, information coming from suppliers (customers) could be more relevant for importing (exporting) due to their upstream (downstream) position in the network.}   While \cite{dhyne2023export} focuses on learning from customers only, we find significant information transmission from both suppliers and customers. Methodologically, \cite{dhyne2023export} employs a similar identification strategy to ours: a combination of multidimensional fixed effects and instruments leveraging the network structure. However, their analysis does not account for geographical and industry-level spillovers. We show that neglecting these factors can lead to overestimating downstream and upstream effects. Finally, we provide explicit conditions under which the previous importing behaviour of distant network connections can serve as a valid instrument to address issues of correlated effects, an aspect absent in \cite{dhyne2023export}. Our approach and findings, therefore,  complement those in \cite{dhyne2023export}.

%Our approach and findings, therefore,  complement those in \cite{bisztray2018learning} and \cite{dhyne2023export}.

In a nutshell, our contribution to the literature can be summarized as follows: we study yet unexplored \textit{firm-level} sources of \textit{heterogeneity} in importing arising \textit{from production network spillovers}. 
%In sum, we contribute to the literature by studying yet unexplored firm-level sources of heterogeneity in importing arising from production network spillovers.
Thanks to the richness of the data at our disposal, we go beyond the spatial and the aggregate input-output dimensions of spillovers. To the best of our knowledge, we are the first to provide empirical evidence that information relevant to importing propagates through supplier-buyer connections in a domestic production network. We show that both upstream and downstream propagation are significant, and we also document important heterogeneities in such propagation effects. 

%The paper is organized as follows. In section \ref{sec:data}, a description of the data is provided. presents the dataset and provides descriptive statistics. Section \ref{strategy} explains the empirical strategy and Section \ref{results} reports the results. Section \ref{s: discussion} summarizes the findings and discusses both the interpretation and limitations of the analysis.

%The rest of the paper is structured as follows. In Section \ref{sec:data} we describe the data. Section \ref{sec:EmpiricalSpecification} lays out our empirical strategy. In sections \ref{sec:Results} and \ref{ss:Heterogenity}, we present the results. Finally, Section \ref{sec:Conclusion} summarizes the findings and discusses possible future avenues of research. Tables displaying the results are in the Appendix.

%Subsection \ref{s: don_syst} describes the Italian system for blood donations, while Subsection \ref{s: avis} introduces the specific characteristics of AVIS and its award scheme. 

\section{Data}
\label{sec:data}
%\khnote{Put this section after the model and before the empirical strategy?}

Spanish businesses and individuals operating as professionals are required to adhere to the Value Added Tax (VAT) regulations. As part of their yearly tax reporting to the Spanish tax authority (Agencia Estatal de Administraci\' {o}n Tributaria, AEAT), they disclose all financial transactions with third parties that exceed a total of 3,005 Euro annually, using the M.347 form.\footnote{More information available at: https://www.agenciatributaria.gob.es. Some firms also report values that are below 3005 Euro.} We have access to this \textit{confidential} dataset of all firm-to-firm transactions subject to VAT from 2010 to 2014. % and use them to reconstruct the production network of Spanish firms.

 While the VAT data we have access to is anonymized, we have information on some important firms' characteristics: type of legal entity, sales, number of employees, industry, labour costs, location at the zip code, and the annual value of trade flows (import and export) with EU countries and with Non-EU Countries. We focus on firms classified as corporations (NIF  type code A)\footnote{Tax Identification Number (NIF) in Spain is the alpha-numeric ID needed by an individual or a legal entity to do any procedures that may have any relevance for the Spanish Tax Agency. We observe only the first character of the NIF (NIF type code), which enables us to distinguish between different types of legal entities.} and limited liability corporations  (LLCs, NIF type code B). We also exclude financial sector firms from the data because of the idiosyncrasies of the financial sector.\footnote{Concretely, we exclude financial firms that, according to the IAE (Impuestos sobre Actividades Economicas) classification, are classified as (a) instituciones financieras, (b) seguros,  (c) auxiliares financieros y seguros, (d) actividades inmobiliares.} Using the  VAT data we construct the production network of Spanish firms. This network is directed and consists of nodes representing firms. We say a connection $j \rightarrow i$ exists between supplier firm $j$ and customer firm $i$ in year $t$  if  $j$ sells intermediate input to $i$ in that year.

%Therefore, we observe the population of Spanish firms not belonging to the financial sector. 
Our initial dataset consists of, on average, 831,525 firms observed from 2010 to 2014. We consider only firms active in all years from 2011--2014, which results in a balanced panel of  611,996 firms, with an average of 8,671,919 yearly connections. We then restrict to stable connections, focusing on those that persist each year throughout 2011-2014, potentially missing in one year only  -- in which case we impute this link for the missing year. This reduces the number of firms to 575,896 (5.9\% reduction). Such a restriction to stable links reduces by 41\%  the sample of links being considered, while the eliminated links account for only 15.8\% of the overall firm-to-firm trade.\footnote{For firms that do not appear in 2010 but appear in all years 2011--2014, we impute the value of the variable from 2011. The results are very similar if we exclude 2010 from our analysis.}

Our focus on stable links is motivated by two key considerations.\footnote{Results obtained without applying this restriction for the population of firms belonging to the Spanish region Valencian Community confirm the main findings of the paper.} 
First, we believe that repeated interactions between trade partners, resulting in largely persistent connections, are more likely to facilitate information transmission and generate positive cost externalities. In contrast, for suppliers or customers whose interaction is relatively short, there is limited opportunity for learning or synergy. Second, if we allowed for short-lived connections that are formed and destroyed during the period under consideration, this would raise issues related to network endogeneity in our estimation. To abstract from these issues, we focus on a persistent network structure, thus ensuring no changes in the network topology during the entire period of our analysis. In the end, our sample embodies a balanced panel representing a network with 575,896 firms and 5,087,373 annual links for the period 2010--2014. We observe firms in 9,014 different zip codes and 287 IAE industries. We define this sample as our \textit{main sample}.

%%%%%%%%%%%%%%%%%%%%%%%%%

%%%%%%% TO DO: NETWORK VARIABLES

% {\color{blue}
% \begin{center}
%     *** To be Included. Table with median and averages of:  degree, number of suppliers, and number of customers all with standard deviations. ***
% \end{center}}

%\subsection{Main sample and import starters}
\subsection{Import starters}
\label{sec:datastart}

%\noindent Since the outcome of interest is \textit{starting to import}, we conduct our analysis on the sample of potential import starters, i.e. firms that have not yet imported from any given source region or country. Therefore, following \cite{bisztray2018learning}, our \textit{analysis sample} is a three-way panel (firm, the origin of importing, year) including only \textit{potential import starters}. The observations concern the firms in the main sample that have not yet imported from some source region (260,668 firms). 

\noindent Since the outcome of interest is \textit{starting to import}, our \textit{analysis sample} is restricted to \textit{potential import starters}. Specifically, following \cite{bisztray2018learning}, the analysis sample is structured as a three-way panel (firm, import origin, year) that includes observations where a firm in the main sample has not yet imported from the specified source country (EU vs. non-EU) up to the previous year.
As shown in Table \ref{tbl:destab1} for the baseline year 2010, compared with the firms not observed to start sourcing inputs from abroad, import starters tend to be bigger, more productive, and have more suppliers and customers.
In Table \ref{tbl:destab2}, one can also notice that firms that start to import tend to have a higher number and a higher proportion of peers that are importing compared to firms that do not begin importing. 
%\footnote{These figures are obtained for the sample of firms used in our main specification of Table \ref{tab: olss1}. The number of suppliers and customers of firms that have not started to import is 11.2 and 10.8, respectively.} 

%Descriptive statistics are reported in Table \ref{tbl:destab1}. 

%On average, they have 20.7 suppliers and 20.2 customers. %I have excluded this sentence because the problem is that these are averages across all starters irrespective of the destination and, though they are correct (I have checked), they are confusing
%On average, 41.19\% of the suppliers and 40.98\% of the customers of each potential import starter are importing. 

\section{Empirical model}\label{sec:EmpiricalSpecification}
\noindent In this section, we describe our empirical strategy. The outcome of interest is a firm's beginning to import from a specific origin. We aim to examine the impact of the importing experience of the firm's peers in the production network, whether suppliers or customers, on this outcome.  Our central hypothesis is two-fold. First, we posit that successful importing relies on import-specific knowledge. Second, we suppose that firms connected in the production network, besides trading goods, also exchange information relevant to importing. As a result, we expect that companies are more likely to start importing if their peers have importing experience and can, therefore, convey import-relevant knowledge. Furthermore, we also conjecture that this knowledge is likely to be more significant when it pertains to importing from a specific region or country of origin. Finally, as in \cite{bisztray2018learning}, we assume that the effect of such information diffusion comes with a time lag -- that is, it takes time for a firm to utilize such acquired knowledge and start importing. 

Since firms typically interact differently with customers and suppliers, we distinguish between the effects attributable to each peer type, situated downstream (customers) or upstream (suppliers), respectively. 
%On the one hand,  customers may have different incentives to share information that is useful for starting to import. %While a firm's customer might benefit from the firm discovering a more suitable or productive input supplier, the firm's current supplier may not share this benefit, as their inputs could be substituted by those from the new foreign supplier. 
\noindent On the one hand, a firm's customers can be expected to have a strong incentive to share information concerning more suitable or productive input suppliers since they would benefit from potentially cheaper or higher-quality intermediate inputs. On the other hand, the firm's suppliers lack this incentive and may try to withhold such information out of the concern of being replaced by other foreign suppliers. However, it is also intuitive that those considerations may be counterbalanced by the relative relevance of the information stemming from customers and suppliers. The information regarding potential foreign suppliers may be more valuable when it originates from suppliers rather than customers, given their position upstream in the production chain. Overall, therefore, it is unclear whether one should expect forward or backward linkages to be a more important source of information spillovers related to importing opportunities and know-how.

To proceed formally with the analysis of the problem, let us think of the production side of the economy as a directed network among the set $N=\{  1,2,...,n\}$ of firms of the economy. Formally, this network is represented by an adjacency matrix $\Gmat = \left(g_{ji}\right)_{j,i =1}^{n}$, we assume $g_{ji} \in \{0,1\}$, where $g_{ji} = 1$ means that $j$ supplies input to firm $i$.\footnote{We have also considered a specification where the weights reflect the relative importance of suppliers (customers) for a given firm, measured by the value of bilateral trade between them. The results are qualitatively similar. We decided to use uniform weighting because it facilitates the interpretation of the results.}
Then, we start with the following empirical equations, for every $i$ that is a potential import starter in each period $t$ (as defined in Section \ref{sec:data}): 
%
%notation, use - for indegree variables, + for out degree variables
\begin{align}\label{eq:basic_spec}
		%\begin{align*}
  \begin{split}
		 \hspace{-20pt} y_{i,t} =& \ \alpha  +  \beta_{D}  \frac{1}{d_i^{-}} \sum_{j \in \suppliers_i} g_{ji} y_{j,t-1} +  \beta_U \frac{1}{d_i^{+}}\sum_{j \in \customers_i} g_{ij} y_{j,t-1}+ \sum_{k=1}^{K} \gamma^{k}x_{i,t}^{k}  + \\ & \sum_{k=1}^{K} \delta_D^{k} \frac{1}{d_i^{-}} \sum_{j \in \suppliers_i} g_{ji} x_{j,t-1}^k  +\sum_{k=1}^{K}\delta_{U}^{k}  \frac{1}{d_i^{+}}\sum_{j \in \customers_i} g_{ij} x_{j,t-1}^{k} +\text{FE} + \varepsilon_{i,t}.
		\end{split}
\end{align}
where
\begin{itemize}[topsep=-3 pt, itemsep= -5 pt]
    \item $y_{i,t} \in \{0,1\}$, is the indicator for firm $i$ importing at year $t$
    \item $x_{j,t}^k$ stands for the $k$th characteristic of firm $j$ at $t$, with $k=1,2,...,K$ 
    \item  $g_{ji} \in \{0,1\}$ indicates whether $j$ is a supplier of $i$ ($g_{ij}= 1$) or not ($g_{ij}= 0$). 
    \item $d_{i}^{-}$ and $d_{i}^{+}$ denote,  respectively, the indegree and outdegree of firm $i$ in the production network, i.e. the cardinality of its set of suppliers $N_i^-=\{ j\in N: g_{ji}=1\}$ and customers $N_i^+=\{ j\in N: g_{ij}=1\}$.\footnote{
Note that $\frac{1}{d_i^{-}}\sum_{j \in N_{i}^{-}}g_{ji}= \frac{1}{d_i^{+}}\sum_{j \in \customers_i} g_{ij}= 1$. Therefore   $\frac{1}{d_i^{-}} \sum_{j \in \suppliers_i} g_{ji} y_{j,t-1}$ and  $\frac{1}{d_i^{+}}\sum_{j \in \customers_i} g_{ij} y_{j,t-1}$ are, respectively, the shares of suppliers and customers importing at $t-1$. }
\end{itemize}

 The main coefficients to be estimated are $\beta_{D}$ and $\beta_{U}$ that measure the effects of the weighted import experience of the suppliers and customers of $i$ on $ i$'s import status. We refer to $\beta_{D}$ as the \textit{downstream (or supplier) peer effect} since it captures an effect that propagates downstream from suppliers through the network, while $\beta_{U}$ is called the \textit{upstream (or customer) peer effect} as it captures an analogous effect that propagates upstream from customers. The coefficient $\gamma_k$ reflects the effect of a firm´s own characteristic on the outcome, whereas coefficients $\delta_{D}^{k}$ and $\delta_{U}^{k}$ capture the effects of a network-based average of peers' characteristics. As we shall discuss later, we estimate different specifications of \eqref{eq:basic_spec} that rely on alternative combinations of firm, time, location, industry, and origin-of-import fixed effects (FE).

 %As explained in \cite{dhyne2023export}, 
 The empirical specification \eqref{eq:basic_spec} can be, for instance, motivated by a model in which the fixed costs of importing depend on the characteristics and actions of network peers. More specifically, the main assumption is that the entry cost into a foreign market decreases with the relevant importing-related information that can be gathered from network peers. In this respect, it is worth noting that our linear-in-means formulation makes a twin assumption that is common in the literature on network-based learning models (see, e.g., \cite{bala1998learning, acemoglu2010spread,golub2010naive}) -- namely, that peer effects depend on the network-weighted outcome of the peers in a linear fashion. The fact that the relevant peer magnitude is taken to be the \textit{average} outcome of peers is akin to the feature posited by \cite{jackson1996strategic} in the context of scientific collaboration; it reflects the natural idea that the intensity of interaction with each peer (supplier or customer in our case) is likely to decrease with the number of peers.\footnote{Alternatively, one could consider a linear-in-sums model,  where peer effects are captured by the number of importing suppliers and customers. This assumption may be inappropriate for at least three reasons. First, the bigger firms tend to have more suppliers/customers, and therefore, the treatment would mechanically be correlated with the firm's size. Second, as explained in \cite{bramoulle2020peer}, the identification of peer effects in linear-in-sum models implies a correlation between the intensity of treatment and the number of peers that has problematic implications for the identification of peer effects in certain settings. Finally, as already mentioning when referring to \cite{jackson1996strategic}, the intensity of communication with each peer is likely to be decreasing with the number of peers.} Since, as it will be clear from the Section \ref{subsec:Identification}, our identification relies on the inclusion of multidimensional fixed effects, we use the linear probability model.\footnote{See \cite{boucher2020binary} for an in-depth discussion of linear probability models in the context of estimation of peer effects in networks.}

In the context of linear-in-means models, the effect of peers' characteristics (observed and unobserved) is known as \textit{contextual peer effect} ($\delta_{D}^{k}$ and $\delta_{U}^{k}$ in \eqref{eq:basic_spec}), while the effect of peers' outcomes is known as  \textit{endogenous peer effect} ($\beta_{U}$ and $\beta_{D}$ in \eqref{eq:basic_spec} ). When these peer effects are contemporaneous, the simultaneity of the outcomes leads to the well-known reflection problem \citep{manski1993identification}, which impedes the identification of endogenous and contextual peer effects.\footnote{Several studies -- including \citet{bramoulle2009identification}, \citet{deGiorgi2010identification} and \citet{lin2010identifying}  -- characterize the conditions that, when agents interact in a network allow one to tackle the reflection problem in linear-in-means model.}  As already mentioned, based on the idea that information diffusion is expected to take time, we assume a delay in the materialization of peer effects. Specifically, we suppose that it is the previous year's peers' import experience that has an effect on firms' decisions to import. Practically, this makes the reflection problem inconsequential in our case, since it breaks the simultaneity of endogenous peer effects and a firm's decision to import. 

Such a delay, however,  does not resolve another challenge that can compromise the identification of peer effects, i.e. the issue of \textit{correlated effects}. These arise when firms connected in the network experience common shocks or share similar unobserved characteristics, thus leading to potential biases in estimating peer effects. In Section \ref{subsec:Identification}, we discuss two incrementally complex identification strategies designed to address this issue in our setting.

%LEAVE FOR LATER

 % Importantly, with the inclusion of fixed effects, we can control for other types of externalities and unobservables that may be relevant to the outcome and are correlated to the import experience and the characteristics of peers, such as productivity, technological spillovers and common determinants of trade costs at the geographical and sectoral level. At the level of aggregation we use in the analyses,

\subsection{Identification}\label{subsec:Identification}
As explained above, to provide causal estimates of peer effects, we need to address the correlated effects that, for example, may stem from common (unobserved) shocks experienced by connected firms.  
%or contextual peer effects operating via unobservable characteristics. 
 We address this issue by employing two nested strategies that build upon each other. In both of them, we assume that, conditional on the set of controls, firms are randomly connected (i.e. conditionally random peers). This means that firms' unobserved characteristics are uncorrelated with their peers' observed and unobserved characteristics, once we condition on the set of observables and unobservables that we consider. Such an assumption is commonly made in the analysis of peer effects with observational data in cross-sectional settings (see \cite{bramoulle2020peer}). In our analysis, this assumption is less restrictive than in those cases because the panel nature of our data allows us to control for a wide array of unobserved factors.

Our first identification strategy, detailed in Section \ref{sss:Strategy1}, relies on an extensive set of fixed effects to account for variation in the unobserved characteristics of a firm, its suppliers, and its customers. This strategy utilizes import-origin variation in the share of importing neighbors independent of their geographical and sectoral distribution, as well as the firm's own time-varying characteristics and those of its neighbors (excluding import-origin-specific factors) to identify peer effects. In Section \ref{sss:Strategy2}, we extend this approach by leveraging the network structure to address remaining identification challenges that emerge at this granular level of variation.

Finally,  it is worthwhile to reiterate that we fix the network $\Gmat$ (by considering only links that appear in all years 2010-2014), which partially addresses the potential issues due to network endogeneity. 
 
 %This decision serves a dual purpose. Firstly, it enables us to focus on the peer effects coming from \textit{stable} peers. Secondly, it essentially makes the network  predetermined in our analysis, thereby making the conditional randomness assumption more plausible.

%, for instance, \cite{feld2017understanding}, \cite{hoekstra2018peer}, and 

\subsubsection{Strategy 1: Fixed-effect regression}\label{sss:Strategy1}

In our first strategy, we use fixed effects to control for the variation in the unobserved characteristics of a firm, its suppliers, and its customers. In the most strict specifications considered, we control for \textit{firm$\times$year} and \textit{industry$\times$zip\_code$\times$origin$\times$year} fixed effects, and estimate the following version of \eqref{eq:basic_spec}:
\begin{align}\label{eq:regression_strategy1} \tag{\ref{eq:basic_spec}.S1}
		%\begin{align*}
  \begin{split}
		 \hspace{-20pt} y_{ihc,t} = &  \beta_{D}  \frac{1}{d_i^{-}} \sum_{j \in \suppliers_i} g_{ji} y_{jhc,t-1} +  \beta_U \frac{1}{d_i^{+}}\sum_{j \in \customers_i} g_{ij} y_{jhc,t-1} + \sum_{k} \gamma^{k}x_{ihc,t}^{k}  + \\ & \sum_{k} \delta_D^{k} \frac{1}{d_i^{-}} \sum_{j \in \suppliers_i} g_{ji} x_{jhc,t-1}^k  +\sum_{k}\delta_{U}^{k}  \frac{1}{d_i^{+}}\sum_{j \in \customers_i} g_{ij} x_{jhc,t-1}^{k} +\mu_{i,t} + \eta_{hc,t} + \varepsilon_{ihc,t},
		\end{split}
\end{align}
\noindent where index $h$ stands for \textit{industry$\times$zip\_code}, and $c$ denotes the import origin (source). The \textit{firm$\times$year}  fixed effects  $\mu_{i,t}$ control for firm-level time-varying observables and unobservables, whereas $\eta_{hc,t}$ denotes \textit{industry$\times$zip\_code$\times$origin$\times$year} fixed effects.  The inclusion of \textit{firm$\times$year} fixed effects implies that we rely on import origin variation in our estimation of endogenous peer effects (parameters $\beta_D$ and $\beta_U$). Moreover, \textit{firm$\times$year} fixed effects absorb the contextual peer effects of any observable characteristics that are not specific to the import origin. Therefore, the inclusion of these fixed effects addresses the potential concern that firms that are more prone to import tend to be connected due to their observed or unobserved characteristic.  The \textit{industry$\times$zip\_code$\times$origin$\times$year} fixed effects ($\eta_{hc,t}$) control for time-varying import origin-specific variables common to firms located at the same zip code and belonging to the same industry. This very demanding set of fixed effects absorbs common shocks and spillovers at the spatial and industry-specific levels (i.e., technological or demand shocks), including the presence of importing neighbours in the geographic/industry network.\footnote{Compared to both \cite{bisztray2018learning} and \cite{dhyne2023export}, we use a more demanding set of fixed effects. \cite{bisztray2018learning} control for  \textit{firm$\times$year} and \textit{origin$\times$year} fixed effects since, in their setup, the spillovers are location-based. \cite{dhyne2023export}, in their most demanding specification, only include firm$\times$year and export-destination$\times$year fixed effects. As it will be clear from the results reported in Section \ref{sec:Results}, the failure to control for location and industry factors may lead to a substantial bias in estimating the peer effects. }  
Furthermore, we recall that peer effects operate in \eqref{eq:regression_strategy1} with a year lag, which rules out correlated effects from non-persistent (temporary) shocks. %We discuss the results of this strategy in Section \ref{ss:Results_Strategy1}.

%noindent After the inclusion of the discussed rich set of fixed effects, the remaining threats to identification stem from import origin variation that is independent of geographical and industry-level variation. For instance, certain types of firms that have a propensity to import from a given destination tend to form supplier-customer linkages independently of their location and industry.

In \eqref{eq:regression_strategy1},  therefore, the identification of the impact of information spillovers from importing neighbours relies on the import-origin variation in the share of importing neighbours that is independent of their geographical and sectoral distribution, as well as of the firm's (not import-origin specific) own time-varying characteristics and the time-varying characteristics of its neighbours. Consequently, potential residual threats to identification must manifest at this nuanced level of variation. 

%This may be caused by correlated unobserved firm-origin-time specific characteristics due, for instance, common shocks.  In the next subsection, we delineate an instrumental variable strategy devised to address the possible presence of correlated effects at this level of variation. 

\subsubsection{Strategy 2: Network instruments}\label{sss:Strategy2}
The strategy described in the previous section accounts for multiple channels through which correlated effects could bias peer effect estimates. These include geographic and industry-level shocks affecting importing behaviour from a specific origin at a very disaggregated level (about three hundred industries times nine thousand zip codes), as well as potential common unobservable factors—unrelated to the origin of imports—that influence the importing behaviour of a firm and its peers. %that impact peer-importing behavior that are .

There are, however, other considerations that need to be taken into account. For example, suppose that firms differ in their time-varying (unobserved) investments in a technology that requires inputs from a specific region. In such cases, firms with higher investments in such technology will be more likely to import from that region. Consequently, if there are technological spillovers across connected firms,  the estimated peer effects in \eqref{eq:basic_spec} may not represent true endogenous peer effects but rather contextual peer effects driven by technological spillovers. This issue persists even when accounting for the extensive set of fixed effects in \eqref{eq:regression_strategy1} and assuming that uncontrolled unobservables are conditionally random across peers. The underlying problem stems from omitted (unobserved) firm-specific variables that vary over time and import origins, and generate contextual peer effects. In this section, we propose a strategy that also addresses this threat to identification.
{
 }
 
This identification strategy takes advantage of the structure of the production network. The key idea is that, under the assumption of conditionally random peers,
%a firm specific observed and unobserved variables are uncorrelated with the characteristics of its second-order peers (who are indirectly connected to the firm via its peers) after controlling for the firm's own characteristics and those of its direct peers.  Thus, 
we can use second-order peers (who are indirectly connected to the firm via its peers) to instrument the importing behaviour of firms' peers. In the following paragraphs, we discuss the details of this idea and its validity. 
 
For exposition simplicity, suppose that,  in \eqref{eq:regression_strategy1}, there is only one relevant observable characteristic $x_{ihc,t}$.   Moreover,  let us denote with $\bar{x}_{ihc,t}^{L}$ and $\bar{y}_{ihc,t}^{L}$, $L \in \{D,U\}$ the network averages of the observed characteristics and outcomes of the respective network peers of $i$. In particular, we have: $\bar{x}_{ihc,t}^{D}  \equiv \frac{1}{\indeg_i} \sum_{j \in \suppliers_i}g_{ji}x_{jhc,t}$; $\bar{x}_{ihc,t}^{U} \equiv \frac{1}{\outdeg_i} \sum_{j \in \customers_i}g_{ij}x_{jhc,t}$; $\bar{y}_{ihc,t}^{D}  \equiv \frac{1}{\indeg_i} \sum_{j \in \suppliers_i}g_{ji}y_{jhc,t}$; and  $\bar{y}_{ihc,t}^{U} \equiv \frac{1}{\outdeg_i} \sum_{j \in \customers_i}g_{ij}y_{jhc,t}$. Finally, let $u_{ihc,t}$ denote the unobserved firm-specific characteristics,\footnote{There are likely many different relevant firm-level unobserved characteristics that can be represented with vector $\mathbf{u}$. For expositional simplicity, we assume there is only one. It will be clear that our conclusions are not affected by this simplification.}  and, importantly,  allow for contextual peer effects with respect to these unobserved attributes.
We now rewrite \eqref{eq:regression_strategy1} by explicitly including the unobserved characteristics $u_{ihc,t}$ and contextual peer effects with respect to these characteristics, both of which are not observed. 
\begin{align}\label{eq:panel_obs_instruments} \tag{\ref{eq:basic_spec}.S2.1}
\begin{split}
    y_{ihc,t} = & \gamma x_{ihc,t} + \delta_D \bar{x}_{ihc, t-1}^{D}+ \delta_U \bar{x}_{ihc, t-1}^{U} + \beta_D \bar{y}_{ihc, t-1}^{D}+ \beta_U \bar{y}_{ihc, t-1}^{U} + \\
    &\zeta_U u_{ihc,t} + \zeta_D \bar{u}_{ihc, t-1}^{D}+ \zeta_U \bar{u}_{ihc, t-1}^{U} + \mu_{i,t} +  \eta_{hc,t}+ \varepsilon_{ihc,t},
    \end{split}
\end{align}
\noindent with $\E(\varepsilon_{ihc,t}|\vec{x}, \vec{u}, \vec{\mu}, \vec{\eta}, \Gmat)=0$, where boldface variables indicate vectors of corresponding variables included in \eqref{eq:panel_obs_instruments}.\footnote{We may allow for non-zero correlation between firm-specific observables and unobservables ($u_{ihc,t}$ and $x_{ihc,t}$) in \eqref{eq:panel_obs_instruments} as this will not affect the estimates of endogenous peer effects we are interested in.} 
%
%In \eqref{eq:panel_obs_instruments}, the variable denoted by  $\varepsilon_{ihc,t}$ in \eqref{eq:regression_strategy1} is decomposed as $\varepsilon_{ihc,t} = \zeta u_{ihc,t} + \zeta_D \bar{u}_{ihc, t-1}^{D}+ \zeta_U \bar{u}_{ihc, t-1}^{U} + \nu_{ihc,t} $, where $\bar{u}_{ihc, t-1}^{D}$ and $\bar{u}_{ihc, t-1}^{U}$ are defined in a way analogous to how $\bar{x}_{ihc, t-1}^{D}$ and $\bar{x}_{ihc, t-1}^{U}$ were defined before.  
Importantly, \eqref{eq:panel_obs_instruments} includes contextual peer effects concerning unobserved $u_{ihc,t}$, which in turn generates the correlated effects problem discussed at the beginning of Section \ref{sss:Strategy2}. 

To see this, consider a customer of firm $i$ and denote it with $k$. By writing a counterpart of \eqref{eq:panel_obs_instruments} for firm $k$ at $t-1$, it becomes clear that 
%
% \begin{align}\label{eq:panel_obs_instruments_lag}  \tag{\ref{eq:basic_spec}.S2.2}
% \begin{split}
%     y_{k,t-1} = & \alpha + \gamma x_{k,t-1} + \delta_D \bar{x}_{k, t-2}^{D}+ \delta_U \bar{x}_{k, t-2}^{U} + \beta_D \bar{y}_{k, t-2}^{D}+ \beta_U \bar{y}_{k, t-2}^{U} + \\
%     &\zeta u_{k,t-1} + \zeta_D \bar{u}_{k, t-2}^{D}+ \zeta_U \bar{u}_{k, t-2}^{U} + \nu_{i,t-1}.
% \end{split}
% \end{align}
% %
% \noindent From \eqref{eq:panel_obs_instruments_lag} it is clear that
%
$\bar{y}^{U}_{khc, t-1}$  is endogenous as it is correlated with  $ \bar{u}_{ihc, t-1}^{U}$ which is not observed, and therefore omitted in \eqref{eq:regression_strategy1}. The analogous argument holds for $\bar{y}^{D}_{khc,t-1}$.  Hence $\beta_{U}$ and $\beta_{D}$ are not identified due to correlated effects. This is true whenever an unobserved \textit{firm-origin-year} specific variable generates contextual peer effects. Note that the first identification strategy described in Subsection \ref{sss:Strategy1} addresses the issue of correlated effects induced by contextual peer effects due to non-origin specific unobservables since we control for  \textit{firm$\times$year and industry$\times$zip\_code$\times$origin$\times$year} fixed effects.

To address this issue, we leverage the network structure. Let $\bar{\bar{y}}^{U}_{ihc, t-2}$ denote the network average of importing status at $t-2$ from a given destination of the second-order customers (customers of customers) of $i$ that are neither direct suppliers nor customers of $i$. Analogously, define by $\bar{\bar{y}}^{D}_{ihc, t-2}$ the network average of importing status of second-order suppliers of $i$ that are not direct peers of $i$. It follows directly from \eqref{eq:panel_obs_instruments} that these averages will affect $\bar{y}^{U}_{ihc, t-1}$ and $\bar{y}^{D}_{ihc, t-1}$ respectively. 

We now argue that $\bar{\bar{y}}^{U}_{ihc, t-2}$ and $\bar{\bar{y}}^{D}_{ihc, t-2}$ are valid instruments for $\bar{y}^{U}_{ihc, t-1}$ and $\bar{y}^{D}_{ihc, t-1}$ respectively,   under the assumption that $cov(u_{ihc,t},u_{ihc, t-q}|\cond)=0$ for all $q\geq 2$, where $\mathbf{z} \equiv (\vec{x}, \vec{\mu}, \vec{\eta}$).  This assumption imposes \textit{limited persistence} in unobserved characteristics and is required for this strategy to work.\footnote{This condition is crucial but has been overlooked in \cite{dhyne2023export}, which employs a similar strategy to estimate peer effects in exporting.} To see this, consider, for instance, a customer of a customer of firm $i$ that is not a direct peer of $i$ and denote it by $\ell$. %Firm $\ell$ enters in $\bar{\bar{y}}^{U}_{ihc, t-2}$.  
Then write \eqref{eq:panel_obs_instruments} for $\ell$ at $t-2$ as follows:
\begin{align}\label{eq:panel_obs_instruments_lagged}\tag{\ref{eq:basic_spec}.S2.2}
\begin{split}
    y_{\ell hc,t-2} = & \gamma x_{\ell hc,t-2} + \delta_D \bar{x}_{\ell hc, t-3}^{D}+ \delta_U \bar{x}_{\ell hc , t-3}^{U} + \beta_D \bar{y}_{\ell hc, t-3}^{D}+ \beta_U \bar{y}_{\ell hc, t-3}^{U} + \\
    &\zeta u_{\ell hc,t-2} + \zeta_D \bar{u}_{\ell hc , t-3}^{D}+ \zeta_U \bar{u}_{\ell hc, t-3}^{U} + \mu_{\ell,t-2} +  \eta_{hc,t-2}+ \varepsilon_{\ell hc,t-2}.
    \end{split}
\end{align}
It is clear that $y_{\ell hc,t-2}$ is correlated with $\bar{y}^{U}_{ihc, t-1}$.  To show that $\bar{\bar{y}}^{U}_{ihc, t-2}$ satisfies the exclusion restriction, we need to argue that $cov(y_{\ell hc,t-2}, \zeta_U u_{ihc,t} + \zeta_D \bar{u}_{ihc, t-1}^{D}+ \zeta_U \bar{u}_{ihc, t-1}^{U}|\mathbf{z}) =0 $. To simplify notation, define $\nu_{ihc,t} \equiv \zeta_U u_{ihc,t} + \zeta_D \bar{u}_{ihc, t-1}^{D}+ \zeta_U \bar{u}_{ihc, t-1}^{U}$. The conditional random network assumption  directly implies
%
%\begin{equation*}
$cov(\gamma x_{\ell hc, t-2} + \delta_D \bar{x}^{D}_{\ell hc, t-3} + \delta_U \bar{x}^{U}_{\ell hc, t-3} + \zeta u_{\ell hc, t-2},\nu_{ihc,t}|\mathbf{z}) =0.$
%\end{equation*}
Moreover, since by construction the customers of $\ell$ are neither suppliers nor customers of $i$, we have that $cov(\bar{u}^{U}_{\ell hc, t-3}, \nu_{ihc,t} |\mathbf{z}) =0.$ However, because at least one first-order supplier of $\ell$ is a first-order customer of $i$, $y_{\ell hc , t-2}$ is correlated with $\nu_{ihc,t}$ whenever $\bar{u}_{\ell hc, t-3}^{D}$  is correlated with $\bar{u}_{i hc, t-1}^{U}$. With conditionally random peers, this will be the case if and only if $cov(u_{i hc,t},u_{i hc, t-2}|\mathbf{z})\neq0$.
%\footnote{The assumption that the network is conditionally random rules out the correlation of unobservables across firms. Since the network is connected, the outcome of $ i$'s supplier or customer of any order observed at any time period will be correlated with $\varepsilon_{it}$ unless we assume some restriction in the form of $cov(u_{it},u_{i, t-q}|\cond)=0$ for all $q \geq q_0$.If we move "further away" in the network, the required $q_0$ increases.}  
%An analogous argument holds for second-order suppliers of $i$ that are not direct peers of $i$.
Finally, the correlation between $y_{\ell hc , t-2}$ and $\nu_{ihc,t}$ may be due to correlation between $\bar{y}_{\ell hc, t-3}^D$ or  $\bar{y}_{\ell hc, t-3}^U$ with $\nu_{ihc,t}$. Nevertheless, by writing down \eqref{eq:panel_obs_instruments} for the first order suppliers (customers) of firm $\ell$ at $t-3$, it becomes clear that as long as $cov(u_{i hc,t},u_{i hc, t-2}|\cond)=0$, this cannot be the case. This concludes the argument that $\bar{\bar{y}}^{U}_{ihc, t-2}$ is valid instrument for $\bar{y}^{U}_{ihc, t-1}$. The analogous argument can be used to show that $\bar{\bar{y}}^{D}_{ihc, t-2}$ is a valid instrument for $\bar{y}^{D}_{ihc, t-1}$.
%Finally, let us note that one could use $x_{}$

\section{Main Results}\label{sec:Results}
In this section, we present the results of our estimates, which rely on the identification strategies discussed in Sections \ref{sss:Strategy1} and \ref{sss:Strategy2}. All tables are reported in the Appendix. 

%\subsection{Strategy 1}\label{ss:Results_Strategy1}

%\khcomment{Results of strategy 1}
In Table \ref{tab: olss1}, we present the estimates of the peer effects that are based on the strategy outlined in Section \ref{sss:Strategy1} (strategy 1).  We estimate different variations of equation (\ref{eq:regression_strategy1}) by OLS, starting from simpler specifications and progressively increasing the complexity. The dependent variable in the specifications considered in the first five columns is the importing status from origin $c \in \{\text{EU}, \text{outside EU} \}$. The sample used is the set of potential import starters from each given import origin, as explained in Section \ref{sec:datastart}.

 In column (1), we initially control for firm fixed effects and \textit{origin$\times$year} fixed effects (see at the bottom of the tables, \textit{id} and \textit{eu-y}, respectively).  The coefficients of interest are reported in the first two rows of Table \ref{tab: olss1}, where $\bar{y}^{D}_{ihc, t-1}$ and $\bar{y}^{U}_{ihc, t-1}$ denote, for each firm $i$ belonging to industry-zip\_code pair $h$, the network-average importing status of its direct suppliers and customers from origin $c$ (EU or extra-EU) in year $t-1$.
% firm $i$'s (belonging to sector-zip\_code pair $h$) direct suppliers' and customers' network average importing status from origin $c$ (EU or extra-EU) in year $t-1$, respectively. 
 The estimated coefficients in this column suggest that a rise of 10 percentage points (from now onward abbreviated to pp) in the proportion of suppliers (customers) that are importing from the same origin at $t-1$ is associated with a 0.314 pp (0.317 pp) increase in the probability of starting importing from that origin. Given that the unconditional probability to start importing (the baseline) in our sample is 3.6\%, this effect amounts to a probability premium of 8.8\% (8.9\%).%\footnote{A rise of 10 pp in the proportion of suppliers (customers) that are importing is approximately equal to one additional supplier (customer) for an average firm in the sample.}  

 %To put these numbers in context, we note that firms in our sample have, on average, 14.36 suppliers, and 12.20 customers. This means that one more importing supplier (customer) leads to an import probability premium of 6.2\% (7.3\%) in the probability of starting importing for an average firm.\footnote{The median number of customers is 2, and the median number of suppliers is 4.}

In column (2), we additionally control for the time-varying observable characteristics of potential import starters at time $t$ and the contextual peer effects -- the average characteristics of their customers at time $t-1$ and those of their suppliers at time $t-1$. In particular, we consider the number of workers, labour costs, the number of suppliers, the number of customers, intermediate inputs cost, sales to other firms, average sales per customer, labour (revenue) productivity (i.e., sales divided by the number of workers), intermediate input (revenue) productivity (i.e., sales divided by intermediate input costs), and the average salary paid. 
%($\text{\textit{suppliers}}_{i,t-1}$ and $\text{\textit{customers}}_{i,t-1}$, respectively).
The estimated coefficients do not change significantly compared to column (1).

In column (3), we introduce \textit{firm$\times$year} fixed effects, which capture firm-level time-varying unobservables and observables through the variable \textit{id-year}. Note that \textit{firm$\times$year} fixed effects account for the contextual peer effects of observable and unobservable firm-year-specific variables. The estimated downstream and upstream effects are only slightly smaller compared to column (1).

 %In column (4), we control for the presence of importers that are neighbors in the geographic/industry network (i.e., importing firms that belong to the same zip code/industry), which are precisely the variables reflecting spillovers in international trade that have been studied by the received literature on the international trade. 
 
In column (4), we control for the presence of importers that are neighbours in the geographic/industry network (i.e., importing firms that belong to the same zip code/industry), capturing location and industry spillovers. The spillovers of this type received significant attention in the international trade literature (see, for instance, \cite{lopez2010imports,harasztosi2011, bisztray2018learning, bekes2020machine, haller2023importer}). 
%
%on which the previous studies of spillovers in international trade have focused. 
%In Table \ref{tab: olss1} 
Concretely, we consider a specification with the additional regressors: $\text{\textit{prop\_imp\_sec}}_{ic,t-1}$, which denotes the proportion of firms that are importing from origin $c$  at $t-1$ and are in the same industry as firm $i$.; $\text{\textit{prop\_imp\_zip}}_{ic,t-1}$, which indicates the proportion of firms that are importing from origin $c$  at $t-1$ and are located in the same zip code as firm $i$; $\text{\textit{prop\_imp\_sec\_zip}}_{ic,t-1}$, which stands for the proportion of firms that are importing from origin $c$ at $t-1$ and are from the same industry and same zip code as firm $i$. The evidence supports the existence of positive and significant location, industry, and location-industry spillovers. We find that firms in the same zip code or/and industry are likely to interact and share relevant importing-related information through channels other than the production networks. Moreover, it may also happen that shocks that affect importing are correlated within location or/and industry.  
%We find evidence suggesting the existence of positive and significant location, sectoral, and location-sectoral spillovers. Firms in the same zip code likely interact through channels other than the production networks. Firms in the same sector may share information about potential suppliers as they use similar production technology. Moreover, shocks that affect importing are likely correlated within the location and sector. Incorporating sectoral and geographical spillovers reduces the estimated peer effects by more than half, underscoring the significance of accounting for these dimensions.  
These findings highlight the necessity of integrating a more nuanced set of fixed effects into the econometric model to capture such localized and industry-specific spillovers as well as absorb possible common unobserved shocks operating at those levels. 
%This refinement is implemented in our next specification, whose results are presented in column (5).
%This is consistent with the existence of location and sector-specific homophily in the production network.

A refinement of our analysis that addresses this concern is conducted through the specification considered in column (5) 
of Table \ref{tab: olss1}. It is our most demanding specification, in that it accounts for time-varying import-origin-specific observables and unobservables common to firms belonging to the same zip code and/or industry. Those fixed effects, captured by the variable denoted by \textit{eu-s-z-y}, also absorb the information regarding the presence of importing neighbours in the geographic/industry network in a nonparametric way. The estimated effects are similar but slightly smaller than those reported in column (4). According to these new estimates, an increase of 10 pp in the share of suppliers (customers) importing leads to a 0.118 pp (0.102 pp) increase in the probability of starting importing from a given area. This equals a probability premium of approximately 3\% calculated at baseline. 
%We present the results of our most demanding and preferred specification in column (5), where we account for time-varying import origin-specific observables and unobservables common to firms belonging to the same zip code and sector. This very demanding set of fixed effects (see at the bottom of the tables, \textit{eu-s-z-y}) also absorbs the information regarding the presence of importing neighbors in the geographic/sectoral network in a nonparametric way. The estimated effects are similar but slightly smaller compared to those reported in column (4). According to these estimates, an increase of 10 pp in the share of suppliers (customers) importing leads to a 0.118 pp (0.102 pp) increase in the probability of starting importing from a given area. This equals a probability premium of 3.32\% (3\%) calculated at the baseline.  

Finally, in column (6), we study a specification that, for every firm and year, focuses on whether the firm in question starts to import in that year, independently of the country of origin. 
% Finally, in column (6) we consider just one observation per firm and year, and select a sample of potential import starters (i.e., observations in which a firm has not yet imported from any destination up until the previous year) to study a different outcome: starting to import, independent of the country of origin. 
In this context, we cannot use \textit{firm$\times$year} fixed effects, so we control for firm fixed effects. We also control for firm-specific variables (at every time $t$) and the associated contextual peer effects  (at  $t-1$) with respect to the same firm characteristics used in column (2) of  Table \ref{tab: olss1}. Additionally, by including \textit{industry$\times$zip\_code$\times$year} fixed effects, we wash out time-varying observables and unobservables common to firms belonging to the same zip code and industry. In this setup, our findings do not support the presence of peer effects in importing. We interpret this result as indicating that spillover effects are primarily significant for knowledge specific to a particular geographical region. The absence of significant spillovers at this level of aggregation is consistent with the findings of \cite{harasztosi2011}, who studies import spillovers of peers located in the same NUTS4 level micro-region.

% \noindent Finally, in column (7) we combine \textit{id-year} and \textit{eu-s-z-y} with the average characteristics of the contacts of firm $i$ that are importing from country $c$ at time $t-1$ ($\text{\textit{ln\_import\_provY}}_{ic,t-1}$, $\text{\textit{ln\_import\_custY}}_{ic,t-1}$, $\text{\textit{ln\_import\_provP}}_{ic,t-1}$, $\text{\textit{ln\_import\_custP}}_{ic,t-1}$).

%\input{MainTables/Table_Strategy1_No3and7}

%\subsection{Strategy 2}\label{ss:Results_Strategy2}
Although in Table \ref{tab: olss1} we control for the observables and unobservables of firms and neighbours at the import\_origin-industry-firm\_location level, it may still be that the productivity gain or cost of importing from a given origin tends to be correlated across neighbours. As we explained in Section \ref{sss:Strategy2}, this may be the case even when the network is conditionally random due to contextual peer effects operating through the unobserved characteristics. 
To tackle this problem, we rely on the instrumental-variable approach described in Section \ref{sss:Strategy2} that uses the importing status of second-order neighbours that are not first-order neighbours as an instrument for peers' importing.\footnote{In the empirical implementation when calculating $\bar{\bar{y}}^{U}_{ihc, t-2}$ we also exclude suppliers of customers of $i$, customers of suppliers of $i$, and second-order suppliers of $i$. While this is not necessary for the identification strategy to work, it facilitates the interpretation of IV estimates as the upstream propagation effect. We use an analogous approach to calculate $\bar{\bar{y}}^{D}_{ihc, t-2}$.} More specifically, we apply this IV approach to the specification estimated in column (5) of Table \ref{tab: olss1}, arriving at the estimation results in Table \ref{tab: IV_fixed_restrictive}.

In column (1)  of Table 2, we instrument variables  $\bar{y}^{D}_{ihc, t-1}$ and $\bar{y}^{U}_{ihc, t-1}$ with $\bar{\bar{y}}^{D}_{ihc, t-2}$ and $\bar{\bar{y}}^{U}_{ihc, t-2}$, which denote the averages of the importing status of the second order suppliers and customers that are not direct suppliers or customers, respectively. The estimated effects, both downstream and upstream, are noticeably larger than those estimated in column (5) of Table 1. According to these estimates, an increase of 10 pp in the share of suppliers (customers) importing leads to an increase in the probability of importing by 0.386 pp (0.684 pp). This translates to approximately a 10.7\% (19.2\%) probability premium at the baseline. An average firm in our sample has 6.9 suppliers and 6.5 customers. This means that having one more supplier (customer) importing from a given area implies an increase in the probability of starting importing from that area by 0.56 pp (1.05 pp). For comparison's sake, we note that the difference in the observed probability to start importing for observations below and above the median of the number of employees is 1.6 pp. Therefore, we conclude that the estimated effects are economically sizable.\footnote{We note that this IV strategy implies a reduction in our sample size, which happens for two reasons. First, since we use variables at $t-2$ as instruments, we reduce the number of years we use in the estimation by 1 (3 instead of 4). Second, we restrict ourselves to firms with both second-order suppliers and second-order customers that are not first-order connections, reducing the sample size even further.} 
%For comparison's sake, we note that moving from the third to fourth quintile in the size distribution implies an increase in the probability of starting importing by 3 pp (see Table \ref{tbl:quintile_size_import}).

As explained in the Section \ref{sec:Intro}, 
%(see also Section \ref{sec:EmpiricalSpecification})
one might expect that the upstream and the downstream effects derived from the possibility of sharing import-relevant information could be of different magnitude, given that the incentives in each case are intuitively quite different. That is, they appear to be strong upstream when customers face the decision of whether to share import-relevant information, while the reciprocal incentives for suppliers to share downstream that information should be quite weaker. The comparison, however, is reversed if we focus instead on the relevance of the information being shared. In this respect, it is intuitive that the information is likely to be more useful when suppliers share it downstream as opposed to when buyers share it upstream. 

In the end, since the aforementioned considerations -- one bearing on \textit{incentives} and the other on \textit{relevance} -- favour a different network direction, it is \textit{a priori} unclear whether upstream or downstream effects should prove to be more consequential. 
%while upstream propagation of importing information from a customer to a supplier may allow the latter to provide the former with a cheaper or better product a favoring the relevance of the downstream channel, downstream information on importing opportunities arriving from the supplier could lead the customer to switching away, totally or partially, from that supplier. s  to a firm's customers may gain from the firm discovering a more suitable or productive input supplier, while the firm's current suppliers may not share this benefit, as those from the new foreign supplier could substitute the inputs they provide. On the other hand, favoring the relevance of the upstream channel, the significance of the information regarding potential foreign suppliers may be more pronounced when it comes from suppliers rather than customers, given their upstream position in the production chain. 
In fact, building upon the estimates displayed in Table \ref{tab: IV_fixed_restrictive}, we find that those two opposing considerations largely compensate for each other. For, even though the point estimate of the upstream effect is larger than the downstream one, these effects are not significantly different at conventional significance levels. We may conclude, therefore, that the evidence does not single either upstream or downstream effects as most effective.

%This is is consistent with the estimates in the literature suggesting that the intermediate inputs are substitutes (for instance, \cite{carvalho2021supply,huremovic2023production}). 

% We note that this IV strategy implies a reduction in our sample size, which happens for two reasons. First, since we use variables at $t-2$ as instruments, we reduce the number of years we use in the estimation by 1 (3 instead of 4). Second, we restrict ourselves to firms with both second-order suppliers and second-order customers that are not first-order connections, reducing the sample size even further.  

% At the bottom of Table 3 we report the under-identification (Kleibergen-Paap LM), weak identification (Kleibergen-Paap Wald F) and over-identifying restriction (Hansen J) statistics and p-values. The statistics of the first two tests indicate that our instruments have predictive power, and the Hansen test suggests that the instruments are identifying the same set of parameters.

Next, in column (2), we enrich the set of instruments by adding to them the network averages of the second-order neighbours importing at $t-3$, i.e. $\bar{\bar{y}}^{D}_{ihc, t-3}$ and $\bar{\bar{y}}^{U}_{ihc, t-3}$. It is clear that these instruments satisfy the exclusion restriction whenever $\bar{\bar{y}}^{D}_{ihc, t-2}$ and $\bar{\bar{y}}^{U}_{ihc, t-2}$ satisfy the exclusion restriction. The estimated effects are now larger but comparable to those from column (1). By including those additional instruments, we can test the over-identifying restriction.  The corresponding test's statistics and p-value (Hansen J) are reported in rows labelled with  $j$ and $jp$.

%\newpage

Finally, in columns (3) and (4), we apply the same instrumental variables strategy used in columns (1) and (2), but without differentiating in terms of the origin of imports.  Thus, as in column (6) in Table \ref{tab: olss1}, we use only one observation per firm-year, and the outcome of interest is starting to import, independently of the origin. In this case, we cannot anymore control for the \textit{firm$\times$year }fixed effects, as they would absorb all the variation in the outcome variable.  Therefore, we control for firm fixed effects (together with \textit{industry$\times$zip\_code$\times$year} fixed effects). We also control for firm-specific variables and the associated contextual peer effects, relying on the same firm characteristics used in column (2) of Table \ref{tab: olss1}. Confirming the results of column (6) in Table \ref{tab: olss1},  we do not find compelling evidence in favour of peer effects in importing. Thus, spillovers still appear relevant only for more specialized information specific to a particular geographical area.

\section{Effect Heterogeneity}\label{ss:Heterogenity}

This section examines various factors driving the possible heterogeneity of the identified spillover effects using a modified version of equation \eqref{eq:regression_strategy1}. In particular, we will allow for the possibility of effect heterogeneity by changing the definition of the treatment variables starting from the specification employed in column (5) of Table \ref{tab: olss1}, which is our most demanding OLS specification, since,  in addition to firm-year fixed effects, it also accounts for time-varying import-origin-specific observables and unobservables common to firms belonging to the same zip code and/or industry. 

We begin by exploring the role of node-level heterogeneity by separately focusing on firm and peer characteristics. We consider effect heterogeneity related to size (i.e., the number of employees), labour productivity (i.e., the ratio between total sales and the number of workers), and connectivity (i.e., the number of suppliers and customers). Furthermore, given their role in linking domestic companies to international markets, we also consider whether the network effects differ between wholesalers and other firms.\footnote{We define wholesalers as firms in NACE industries 45, 46, and 47. The importance of these trade intermediaries, both at the export and the import side, has been documented by \cite{bernard2010wholesalers} and \cite{grazzi2016indirect}.}  %The focus on wholesalers is motivated by the findings in \cite{dhyne2023export}, which highlight notable idiosyncrasies of wholesaler firms regarding peer effects in export activities.

Next, we analyze link-level heterogeneity, investigating the potential effects of factors specific to the match between firms and their suppliers or customers. First, we study possible complementarities between firm and peer characteristics, considering each of the characteristics mentioned above. Second, we investigate whether spillovers vary based on firms being in the same industry or location or having a reciprocal relationship as both supplier and customer.

\subsection{Firm characteristics}\label{ss:HetFirm}

{%We consider heterogeneity concerning firm size, labor productivity (the ratio between total sales and the number of workers), intermediate input productivity (the ratio between total sales and the number of domestic suppliers),  and connectivity (number of suppliers and customers). Furthermore, given the specific role of trade intermediaries, we also consider whether the network effects differ between wholesalers and other firms.\footnote{We define wholesalers as firms in NACE sectors 45, 46, and 47.  The focus on wholesalers is motivated by the findings in \cite{dhyne2023export}, which highlight notable idiosyncrasies of wholesaler firms regarding peer effects in export activities.}
%On the one hand, wholesalers involved in importing activities are likely to be better informed about foreign markets compared to other importers, owing to their specialization in resale and their role in connecting different economic agents. At the same time, they may also have a stronger incentive to protect this knowledge to avoid losing domestic clients.
%\footnote{While other importers also have an interest in safeguarding their market knowledge from their customers, this is particularly relevant for wholesalers, as their business model relies on reselling imported products, which makes their offerings more easily substitutable.} 
%Wholesalers that are not importing could be expected to be particularly interested in learning additional trading opportunities that can be used to resell imported products to non-importing firms.
}

% We single out wholesalers for two reasons. know-how about the feasibility of possible matches between specific foreign providers and domestic buyers
We start by estimating \textit{heterogeneous effects associated with firm characteristics}. To construct the variables that will be interacted with the treatment variables to capture treatment effect heterogeneity based on firm observables, we proceed as follows. For each firm, we fix the values of the characteristics to those observed in 2010. For each firm characteristic, except for wholesaler status, we divide the firms into two groups using the median of the empirical distribution as the cutoff: a \textit{low} group comprising firms with values below the median and a \textit{high} group comprising firms with values above the median.\footnote{Our results remain robust when using the third quartile as the cutoff instead of the median.} For wholesaler status, we differentiate firms directly into two categories: wholesalers and non-wholesalers.

As stated above, our estimation approach builds upon the specification considered in column (5) of Table \ref{tab: olss1}. A modification of such specification is applied \textit{separately} to each characteristic as follows:\footnote{Since in this specification the contextual peer effects, as well as the firm-specific observables, are absorbed with \textit{firm $\times$ year} fixed effects, we do not write them explicitly in the empirical models in this section. We do not observe \textit{firm-year-import\_origin} specific variables.}
\begin{align}\label{eq:regression_het_firm} 
\tag{\ref{eq:basic_spec}.H.1}
  \begin{split}
     \hspace{-20pt} y_{ihc,t} =   \sum_{v \in \{Low,High\}} \left( \beta_{D}^{v} z_{i}^{v} \bar{y}^D_{ihc,t-1} + \beta_U^{v} z_{i}^{v} \bar{y}^{U}_{ihc,t-1} \right) + \mu_{i,t} + \eta_{hc,t} + \varepsilon_{ihc,t}.
  \end{split}
\end{align}
In the equation above, $z_{i}^{v}$ denotes the binary variables that, for each characteristic under consideration, indicate whether firm $i$ exhibits a \textit{High} value (above the median) or a \textit{Low} value (below the median). This approach allows the effect of peer spillovers to vary based on whether the firm’s characteristic lies above or below the median. The results are presented in Table \ref{tbl:het_firm_char}.

%
%\begin{align}\label{eq:regression_het_firm} \tag{\ref{eq:basic_spec}.H.1}
		%  \begin{split}
		% \hspace{-20pt} y_{ihc,t} =   \beta_{D}^{\ell} z_{i,t}^{\ell} \bar{y}^D_{ihc,t-1} + \beta_U^{\ell} z_{i,t}^{\ell}\bar{y}^{U}_{ihc,t-1} + \mu_{i,t} + \eta_{hc,t} + \varepsilon_{ihc,t}.
%		\end{split}
%\end{align}

%
%\input{MainTables/Table_Het_FirmChar}
%

We find that larger firms (measured by the number of workers and the number of suppliers/customers), as well as the more productive ones, utilize the knowledge about importing acquired from peers more effectively. These results are aligned with those obtained by the literature on location spillovers in imports \citep{bisztray2018learning, bekes2020machine, haller2023importer} but contrast with the findings of \cite{dhyne2023export} in the context of production network spillovers in exports. Our results may be understood as reflecting economies of scale in learning how to import or in implementing it.\footnote{\cite{lu2024firms} propose a dynamic model of importing where such economies of scale arise.}
%Firm cthat are already more productive, importing high-quality or more specialized inputs boosts their efficiency relatively more. These firms have technologies or processes that are particularly well-suited to benefit from a greater variety and higher quality of inputs. Second, for firms that produce and sell more, even small reductions in their input costs (through importing cheaper or better inputs) have a larger impact on their overall profitability - which may be due to techn.
Finally, we find that wholesalers are more likely to respond to import knowledge from their peers than non-wholesaler firms.\footnote{These differences in peer effects based on firm characteristics are both economically sizable and statistically significant at the one percent level.} This finding is consistent with the hypothesis that wholesalers are especially interested in expanding their trading opportunities because they can exploit their role as intermediaries for non-importing firms.\footnote{Note that, as a further source of contrast with \cite{dhyne2023export}, our findings show that not only wholesalers learn from both customers and suppliers, but so do other firms. Their analysis concludes that non-wholesalers learn exclusively from customers.}
%As suggested,

\subsection{Peers characteristics}\label{ss:HetPeers}

Next, we address the estimation of \textit{heterogeneous effects derived from customer and supplier characteristics}. To this end, we rely on the following regression:
\begin{align}\label{eq:regression_het_peer} \tag{\ref{eq:basic_spec}.H.2}
  \begin{split}
     \hspace{-20pt} y_{ihc,t} =   \sum_{v \in \{Low,High\}} \left( \beta_{D}^{v}  \bar{y}^{D, v}_{ihc,t-1} + \beta_U^{v} \bar{y}^{U,v}_{ihc,t-1} \right) + \mu_{i,t} + \eta_{hc,t} + \varepsilon_{ihc,t},
  \end{split}
\end{align}

%
%\begin{align}\label{eq:regression_het_peer} \tag{\ref{eq:basic_spec}.H.2}
%  \begin{split}
%		 \hspace{-20pt} y_{ihc,t} =   \beta_{D}^{\ell}  \bar{y}^{D, \ell}_{ihc,t-1} + \beta_U^{\ell} \bar{y}^{U,\ell}_{ihc,t-1} + \mu_{i,t} + \eta_{hc,t} + \varepsilon_{ihc,t},
%		\end{split}
%\end{align}
%
\noindent where $\bar{y}^{D, v}_{ihc,t-1}$ and  $\bar{y}^{U, v}_{ihc,t-1}$  for a given firm $i$ denote the number of importing suppliers and customers of firm $i$ in category $v \in \{Low, High\}$ (i.e.,
with the value of the considered characteristic being lower or higher than the median, respectively) divided by the total number of suppliers and customers of firm  $i$, respectively. Therefore, the sum of the value of these variables computed by category at the supplier (customer) side equals the value of the downstream (upstream) spillover variable used in the main analysis.\footnote{The weighted average of the estimated effects, according to the share of importing contacts in each category, gives the aggregate effects reported in column (5) of Table \ref{tab: olss1}.} This approach allows the effect of peer spillovers to vary depending on whether the spillovers are associated with peers whose characteristics lie above or below the median. The results are reported in Table \ref{tbl:het_peer_char}.

%with those obtained by the literature on location spillovers in imports \citep{bisztray2018learning, bekes2020machine, haller2023importer}
%This contrasts with the findings of \cite{bisztray2018learning}, which in the context of location and managerial networks, arrives at the opposite conclusion. 
We find that the spillovers, in general, tend to be stronger when coming from smaller and less productive firms. That is, big and more productive firms are relatively worse at disseminating information about importing (or are more capable of shielding such information). The only exception is the labour productivity of customers, for which we do not find a statistically significant difference between the strength of spillovers of Low and High peers. This general pattern contrasts with the findings obtained by the literature on location spillovers in imports \citep{bisztray2018learning, bekes2020machine, haller2023importer} that arrive at the opposite conclusion.
%comment on wholesalers....
Regarding the role of wholesalers as peers, it is important to consider that a firm sourcing intermediate inputs from an importing wholesaler might reflect a pre-existing intent to import those inputs directly, hindered by barriers such as fixed costs or uncertainty about the quality of imported goods. Consequently, estimating relatively stronger downstream import spillovers when providers are wholesalers could indicate that this spurious mechanism is contaminating the significant downstream effects estimated in Section \ref{sec:Results}. However, the results in the last column of Table \ref{tbl:het_peer_char} show that firms do not learn from suppliers that are wholesalers, while they do learn from both wholesaler and non-wholesaler customers, thus effectively ruling out the aforementioned spurious mechanism.

\subsection{Firm and Peers characteristics}\label{ss:HetLink}

%The spillovers from peer 
%(i.e. productivity and size)
In this paragraph, we jointly consider the role of firms' and peers' characteristics in explaining effect heterogeneity. 
We begin by examining how the heterogeneity in peer effects, as identified in Table \ref{tbl:het_firm_char} and Table \ref{tbl:het_peer_char}, is influenced by the characteristics of both the firm and its peers. We then investigate how the spillovers vary depending on whether they occur between firms within the same industry, within the same province or zip-code area, and through reciprocal forward and backward linkages. 

%
%We turn to explore how firm and peer characteristics interact in  shaping diffusion. Interaction effects are potentially important because  their presence indicates that sorting firms can further increase the adoption of good business practices.

To explore how the characteristics considered in Tables \ref{tbl:het_firm_char} and \ref{tbl:het_peer_char} interact we  estimate the following specification: %that employs as treatment variables all the possible interactions between the $z_{i}^{v}$ variables and the $\bar{y}^{S, v}_{ihc,t-1}$ variables (with $v \in \{L,H\}$ and $S \in \{D,U\}$):
\begin{align}
\label{eq:regression_het_firm_peer}
\tag{\ref{eq:basic_spec}.H.3}
y_{ihc,t} = & \sum_{S \in \{D,U\}} \left( \beta_{S}^{L,L} \bar{y}^{S}_{ihc,t-1} + \beta_S^{H,L}  \bar{y}^{S}_{ihc,t-1} z_{i}^{H} \right. \notag \\ 
& \quad \left. + \beta_{S}^{L,H} \bar{y}^{S,H}_{ihc,t-1} + \beta_S^{H,H}  \bar{y}^{S,H}_{ihc,t-1} z_{i}^{H} \right) + \mu_{i,t} + \eta_{hc,t} + \varepsilon_{ihc,t}, \notag 
\end{align}
\noindent where, as in the previous specifications, $\bar{y}^{D}_{ihc, t-1}$ ($\bar{y}^{U}_{ihc, t-1}$) represents the network-average importing status of its direct suppliers (customers) from origin $c$ (EU or extra-EU) in year $t-1$; $z_{i}^{H}$ denotes the binary variable that indicates whether firm $i$ exhibits a value of the considered characteristic that is above the median; and $\bar{y}^{D, v}_{ihc,t-1}$ ($\bar{y}^{U, v}_{ihc,t-1}$) represents the number of importing suppliers (customers) of firm $i$ in category $v \in \{L,H\}$ divided by the total number of suppliers (customers) of firm  $i$. The estimated coefficients are reported in Table \ref{tbl:het_firmpeer_char}. If we focus on the specification of the first column considering firm size, $\beta_{S}^{L,L}$ measure the spillover effect to a small firm from small importing peers, $\beta_{S}^{H,L}$ and $\beta_{S}^{L,H}$ represent the additional spillover effect when the firm is big and from big importing peers, respectively, and $\beta_{S}^{H,H}$ measures the possible complementarity effect between firm and peers size.

To illustrate how the estimated effects of peer spillovers vary depending on whether both a firm's characteristic and its peers' characteristic fall above or below the median—or whether they lie in opposite parts of the distribution—Table \ref{tbl:het_firmpeer_char_groups} conveniently aggregates the estimated coefficients of \ref{eq:regression_het_firm_peer} reported in Table \ref{tbl:het_firmpeer_char}.
In this way, instead of presenting the results as incremental effects with respect to a baseline category as in Table \ref{tbl:het_firmpeer_char} (i.e., the coefficient associated to $\beta_{S}^{L,L}$ in \ref{eq:regression_het_firm_peer} measures the spillover effect to a small firm from small importing peers; notice that the first and fifth rows of estimated average effects in Table \ref{tbl:het_firmpeer_char} and in Table \ref{tbl:het_firmpeer_char_groups} are the same), \ref{tbl:het_firmpeer_char_groups} shows how the estimated spillover effects differ across the four possible types of links defined by the values (low or high) of the considered characteristic of a firm and its peer.\footnote{The estimates presented in \ref{tbl:het_firmpeer_char_groups} can also be derived using a specification that expresses the treatment effects by interacting the $z_{i,t}^{v}$ variables with the $\bar{y}^{S, v}_{ihc,t-1}$ variables (with $v \in \{L,H\}$ and $S \in \{D,U\}$). }

The results in Table \ref{tbl:het_firmpeer_char_groups} align with those in Tables \ref{tbl:het_firm_char} and \ref{tbl:het_peer_char}, confirming that high-performance firms (i.e., large, well-connected, or highly productive) benefit relatively more from import spillovers, while spillovers originating from high-importing peers tend to be less relevant. However, they also reveal that spillovers are particularly weak—or even absent—when high-performance firms are linked to low-performance importing peers. This pattern suggests that information on importing opportunities from low-performance suppliers may be less relevant or useful for high-performance firms, potentially due to differences in the quality of foreign intermediate inputs used by low- and high-performance firms and the resulting mismatch in their input quality needs.

Regarding the role of wholesalers,
%with respect to downstream spillovers,
it is reassuring to notice that firms that are not wholesalers do not learn from suppliers that are wholesalers but do learn from non-wholesalers. 
%. Looking at upstream spillovers, we find that with respect to the positive and significant baseline spillover effect between a non-wholesaler firm and non-wholesaler customers, the additional effect for a wholesaler firm (wholesalers customers) is positive (negative) and significant, but, differently from downstream spillovers, we do not detect any additional interaction effect. Therefore, concerning the previous results of the specifications that consider separately firm and customer characteristics, 
Finally, we confirm that wholesalers are more likely to respond to import knowledge from their customers than non-wholesaler firms and find new evidence suggesting that (non-wholesaler) firms learn with a lower intensity from wholesaler customers.

%Regarding the role of wholesalers as peers, it is important to consider that a firm sourcing intermediate inputs from an importing wholesaler might reflect a pre-existing intent to import those inputs directly, hindered by barriers such as fixed costs or uncertainty about the quality of imported goods. Consequently, estimating relatively stronger downstream import spillovers when providers are wholesalers could indicate that this spurious mechanism is contaminating the significant downstream effects estimated in Section \ref{sec:Results}. However, the results in the last column of Table \ref{tbl:het_peer_char} show that firms do not learn from suppliers that are wholesalers, while they do learn from both wholesaler and non-wholesaler customers. Effectively, this rules out the aforementioned spurious mechanism. With respect to upstream spillovers, we find that firms learn with the same intensity from wholesalers and other customers.

%Finally, we find that wholesalers are more likely to respond to import knowledge from their peers than non-wholesaler firms

%\input{MainTables/Table_Het_FirmPeer_Char}

%Second, we explore if the spillovers are different for firms that belong to the same industry, are located in the same province or zip-code area, and come from a firm that is both supplier and customer (i.e., a reciprocal relationship).
We now investigate if the spillovers are different when occurring between firms that belong to the same industry, are located in the same province or zip-code area, or are connected through both forward and backward linkages  (i.e., a reciprocal relationship).
To do so,  we estimate the following regression
\begin{align}\label{eq:regression_het_peer} \tag{\ref{eq:basic_spec}.H.4}
  \begin{split}
     \hspace{-20pt} y_{ihc,t} =   \sum_{v \in \{N,Y\}} \left( \beta_{D}^{v}  \bar{y}^{D, v}_{ihc,t-1} + \beta_U^{v} \bar{y}^{U,v}_{ihc,t-1} \right) + \mu_{i,t} + \eta_{hc,t} + \varepsilon_{ihc,t},
  \end{split}
\end{align}
where $\bar{y}^{D, v}_{ihc,t-1}$ and  $\bar{y}^{U, v}_{ihc,t-1}$  for a given firm $i$ denote the number of importing suppliers and customers of firm $i$ in category $v$ divided by the total number of suppliers and customers of firm  $i$, respectively. In \eqref{eq:regression_het_peer}, $v=Y$ ($v=N$) refers to the group of peers that (do not) belong to the same industry or are (not) located in the same zip code or province or (do not) form a reciprocal relationship with firm $i$. The results are reported in Table \ref{tbl:het_link}.  %Therefore, the sum of the value of these variables computed by category at the supplier (customer) side equals the value of the upstream (downstream) spillover variable used in the main analysis (and the weighted average of the estimated effects, according to the share of importing contacts in each category, gives the aggregate effects reported in column 5 of Table \ref{tab: olss1}).

%Let $w_{it}^{\ell}$ denote an indicator taking value $1$ if the firm and its supplier (customer) belong to the same sector ($\ell$) or are located in the same zip code ($\ell$), or form a reciprocal relation. We estimate the following regressions (one per dimension of heterogeneity). 
%
%\begin{align}\label{eq:regression_het_link} \tag{\ref{eq:basic_spec}.H.3}
	
%  \begin{split}
%		 \hspace{-20pt} y_{ihc,t} =  \beta_{D}^{\ell} w_{it}^{\ell} \bar{y}^{D, \ell}_{ihc,t-1} + 
%          \beta_{D} \bar{y}^{D}_{ihc,t-1} +
%         \beta_{U}^{\ell} w_{it}^{\ell} \bar{y}^{U, \ell}_{ihc,t-1} + 
%          \beta_{U} \bar{y}^{U}_{ihc,t-1} +  \mu_{i,t} + \eta_{hc,t} + \varepsilon_{ihc,t}.
%		\end{split}
%\end{align}

We find that spillovers tend to be higher when coming from firms from the same industry, which is intuitive given that those firms are likely to use a similar mix of inputs in production.  The spillovers are also stronger from reciprocal relations, in which both firms buy and sell to each other. Interestingly, spillovers are stronger from peers located in different locations (identified by the zip code) or provinces (Spain has 50 provinces).  Considering that geographic proximity strongly influences the likelihood of connections between firms, this finding evokes the concept of the "strength of weak ties" effect \citep{granovetter1973strength}, emphasizing the importance of non-localized connections in providing access to new information and opportunities. Moreover, it highlights the role of information transmission through supplier-customer linkages over and above the geographic or sector dimensions.

%In \eqref{eq:regression_het_link} $\beta_{D}^{\ell}$ and $\beta_{U}^{\ell}$ measure the differential peer effect coming from firms that belong to the same sector, are located at the same zip-code or are in reciprocal relation. 

%\input{MainTables/Table_Het_Link}
\section{Concluding remarks}\label{sec:Conclusion}
%In this paper, we study an unexplored dimension of firms' international trade behavior associated with their position in the domestic production network. 
In this paper, we examine a previously unexplored aspect of firms' international trade behaviour, focusing on how their position within the domestic production network influences their importing decisions. To do this, we use a rich dataset provided by the Spanish Tax Agency (AEAT), which provides information about the universe of annual firm-to-firm transactions in the period 2010--2014.  Using a combination of different identification strategies, we find evidence that suppliers' and customers' importing significantly affects a firm's decision to start importing from a given geopolitical area. Larger firms are better at taking advantage of information but less effective at disseminating it (more successful in protecting it). Linkages with geographically distant firms provide relatively more useful information to start importing.

%Our identification relies on standard assumptions shared with other papers aiming to estimate peer effects in social networks, and  settings. We assume the network is fixed and random conditional on observables and unobservable characteristics we control for and we use instruments based on the behavior of  . 

Our identification strategy relies on standard assumptions used in the literature on cross-sectional peer effects in social networks and applies insights from this literature to a panel-data production network setting. In this context, we derive new conditions under which the network instruments can be used to address the issue of correlated effects.

The approach presented in this paper offers several promising directions for further research. For instance, the information about import opportunities may propagate through the network beyond the first-order connections, which may lead to an important amplification of the effects we study here. Furthermore, the mechanism we examine is not limited to firms’ import decisions but may also play a role in shaping domestic firm-to-firm connections. We leave studying these important issues for future research.

%\newpage
\bibliographystyle{apalike}
\setlength{\bibsep}{1pt plus 0.3ex}
\bibliography{Bibliography_TradeDiffusion}

\begin{thebibliography}{}

\bibitem[Acemoglu et~al., 2010]{acemoglu2010spread}
Acemoglu, D., Ozdaglar, A., and ParandehGheibi, A. (2010).
\newblock Spread of (mis) information in social networks.
\newblock {\em Games and Economic Behavior}, 70(2):194--227.

\bibitem[Allen, 2014]{allen2014information}
Allen, T. (2014).
\newblock Information frictions in trade.
\newblock {\em Econometrica}, 82(6):2041--2083.

\bibitem[Amiti and Konings, 2007]{amiti2007trade}
Amiti, M. and Konings, J. (2007).
\newblock Trade liberalization, intermediate inputs, and productivity: Evidence
  from indonesia.
\newblock {\em American Economic Review}, 97(5):1611--1638.

\bibitem[Antras et~al., 2017]{antras2017margins}
Antras, P., Fort, T.~C., and Tintelnot, F. (2017).
\newblock The margins of global sourcing: Theory and evidence from us firms.
\newblock {\em American Economic Review}, 107(9):2514--2564.

\bibitem[Bala and Goyal, 1998]{bala1998learning}
Bala, V. and Goyal, S. (1998).
\newblock Learning from neighbours.
\newblock {\em The review of economic studies}, 65(3):595--621.

\bibitem[B{\'e}k{\'e}s and Harasztosi, 2020]{bekes2020machine}
B{\'e}k{\'e}s, G. and Harasztosi, P. (2020).
\newblock Machine imports, technology adoption, and local spillovers.
\newblock {\em Review of World Economics}, 156:343--375.

\bibitem[Bernard et~al., 2010]{bernard2010wholesalers}
Bernard, A.~B., Jensen, J.~B., Redding, S.~J., and Schott, P.~K. (2010).
\newblock Wholesalers and retailers in us trade.
\newblock {\em American Economic Review}, 100(2):408--413.

\bibitem[Bernard et~al., 2009]{Bernard2009}
Bernard, A.~B., Jensen, J.~B., and Schott, P.~K. (2009).
\newblock Importers, exporters, and multinationals: A portrait of firms in the
  u.s. that trade goods.
\newblock In T., D., J.~B., J., and M., R., editors, {\em Producer Dynamics:
  New Evidence from Micro Data}. University of Chicago Press, Chicago.

\bibitem[Bernard and Moxnes, 2018]{bernard2018networks}
Bernard, A.~B. and Moxnes, A. (2018).
\newblock Networks and trade.
\newblock {\em Annual Review of Economics}, 10:65--85.

\bibitem[Bernard et~al., 2019]{bernard2019production}
Bernard, A.~B., Moxnes, A., and Saito, Y.~U. (2019).
\newblock Production networks, geography, and firm performance.
\newblock {\em Journal of Political Economy}, 127(2):639--688.

\bibitem[Bisztray et~al., 2018]{bisztray2018learning}
Bisztray, M., Koren, M., and Szeidl, A. (2018).
\newblock Learning to import from your peers.
\newblock {\em Journal of International Economics}, 115:242--258.

\bibitem[B{\o}ler et~al., 2015]{boler2015r}
B{\o}ler, E.~A., Moxnes, A., and Ulltveit-Moe, K.~H. (2015).
\newblock R\&d, international sourcing, and the joint impact on firm
  performance.
\newblock {\em American Economic Review}, 105(12):3704--3739.

\bibitem[Boucher and Bramoull{\'e}, 2020]{boucher2020binary}
Boucher, V. and Bramoull{\'e}, Y. (2020).
\newblock Binary outcomes and linear interactions.

\bibitem[Bramoull{\'e} et~al., 2009]{bramoulle2009identification}
Bramoull{\'e}, Y., Djebbari, H., and Fortin, B. (2009).
\newblock Identification of peer effects through social networks.
\newblock {\em Journal of econometrics}, 150(1):41--55.

\bibitem[Bramoull{\'e} et~al., 2020]{bramoulle2020peer}
Bramoull{\'e}, Y., Djebbari, H., and Fortin, B. (2020).
\newblock Peer effects in networks: A survey.
\newblock {\em Annual Review of Economics}, 12:603--629.

\bibitem[Carvalho et~al., 2021]{carvalho2021supply}
Carvalho, V.~M., Nirei, M., Saito, Y.~U., and Tahbaz-Salehi, A. (2021).
\newblock Supply chain disruptions: Evidence from the great east japan
  earthquake.
\newblock {\em The Quarterly Journal of Economics}, 136(2):1255--1321.

\bibitem[Carvalho and Voigtl{\"a}nder, 2014]{carvalho2014input}
Carvalho, V.~M. and Voigtl{\"a}nder, N. (2014).
\newblock Input diffusion and the evolution of production networks.
\newblock Technical report, National Bureau of Economic Research.

\bibitem[Castellani et~al., 2010]{castellani2010firms}
Castellani, D., Serti, F., and Tomasi, C. (2010).
\newblock Firms in international trade: Importers’ and exporters’
  heterogeneity in italian manufacturing industry.
\newblock {\em The World Economy}, 33(3):424--457.

\bibitem[Chaney, 2016]{Chaney2016}
Chaney, T. (2016).
\newblock Networks in international trade.
\newblock In Yann, B., Galeotti, A., and Rogers, B., editors, {\em The Oxford
  Handbook of the Economics of Networks}. Oxford University Press USA.

\bibitem[Chaney, 2018]{chaney2018gravity}
Chaney, T. (2018).
\newblock The gravity equation in international trade: An explanation.
\newblock {\em Journal of Political Economy}, 126(1):000--000.

\bibitem[Coe and Helpman, 1995]{coe1995international}
Coe, D.~T. and Helpman, E. (1995).
\newblock International r\&d spillovers.
\newblock {\em European economic review}, 39(5):859--887.

\bibitem[Colantone and Crin{\`o}, 2014]{colantone2014new}
Colantone, I. and Crin{\`o}, R. (2014).
\newblock New imported inputs, new domestic products.
\newblock {\em Journal of International Economics}, 92(1):147--165.

\bibitem[Corcos and Haller, 2023]{haller2023importer}
Corcos, G. and Haller, S. (2023).
\newblock Importer dynamics: Do peers matter?
\newblock {\em CESifo Working Paper}.

\bibitem[De~Giorgi et~al., 2010]{deGiorgi2010identification}
De~Giorgi, G., Pellizzari, M., and Redaelli, S. (2010).
\newblock Identification of social interactions through partially overlapping
  peer groups.
\newblock {\em American Economic Journal: Applied Economics}, 2(2):241--275.

\bibitem[Dhyne et~al., 2023]{dhyne2023export}
Dhyne, E., Ludwig, P., and Vandenbussche, H. (2023).
\newblock Export entry and network interactions: Evidence from the belgian
  production network.
\newblock Technical report.

\bibitem[Eaton et~al., 2024]{eaton2024search}
Eaton, J., Eslava, M., Jinkins, D., Krizan, C.~J., and Tybout, J.~R. (2024).
\newblock A search and learning model of export dynamics.
\newblock Technical report, National Bureau of Economic Research.

\bibitem[Eaton et~al., 2022]{eaton2022two}
Eaton, J., Jinkins, D., Tybout, J.~R., and Xu, D. (2022).
\newblock Two-sided search in international markets.
\newblock Technical report, National Bureau of Economic Research.

\bibitem[Eaton et~al., 2011]{eaton2011anatomy}
Eaton, J., Kortum, S., and Kramarz, F. (2011).
\newblock An anatomy of international trade: Evidence from french firms.
\newblock {\em Econometrica}, 79(5):1453--1498.

\bibitem[Goldberg et~al., 2010]{goldberg2010imported}
Goldberg, P.~K., Khandelwal, A.~K., Pavcnik, N., and Topalova, P. (2010).
\newblock Imported intermediate inputs and domestic product growth: Evidence
  from india.
\newblock {\em The Quarterly journal of economics}, 125(4):1727--1767.

\bibitem[Golub and Jackson, 2010]{golub2010naive}
Golub, B. and Jackson, M.~O. (2010).
\newblock Naive learning in social networks and the wisdom of crowds.
\newblock {\em American Economic Journal: Microeconomics}, 2(1):112--149.

\bibitem[Granovetter, 1973]{granovetter1973strength}
Granovetter, M.~S. (1973).
\newblock The strength of weak ties.
\newblock {\em American journal of sociology}, 78(6):1360--1380.

\bibitem[Grazzi and Tomasi, 2016]{grazzi2016indirect}
Grazzi, M. and Tomasi, C. (2016).
\newblock Indirect exporters and importers.
\newblock {\em Review of World Economics}, 152:251--281.

\bibitem[Halpern et~al., 2015]{halpern2015imported}
Halpern, L., Koren, M., and Szeidl, A. (2015).
\newblock Imported inputs and productivity.
\newblock {\em American Economic Review}, 105(12):3660--3703.

\bibitem[Hanwei et~al., 2024]{kalina2024}
Hanwei, H., Kalina, M., Oscar, P., and Pisch, .~F. (2024).
\newblock Firm heterogeneity and imperfect competition in global production
  networks.

\bibitem[Harasztosi, 2011]{harasztosi2011}
Harasztosi, P. (2011).
\newblock Export and import spillovers in hungary.
\newblock {\em Unpublished manuscript}.

\bibitem[Huremovic et~al., 2023]{huremovic2023production}
Huremovic, K., Jim{\'e}nez, G., Moral-Benito, E., Peydr{\'o}, J.-L., and
  Vega-Redondo, F. (2023).
\newblock Production and financial networks in interplay: Crisis evidence from
  supplier-customer and credit registers.
\newblock {\em Available at SSRN 4657236}.

\bibitem[Jackson and Wolinsky, 1996]{jackson1996strategic}
Jackson, M.~O. and Wolinsky, A. (1996).
\newblock A strategic model of social and economic networks.
\newblock {\em journal of economic theory}, 71:44--74.

\bibitem[Johnson and Noguera, 2017]{johnson2017portrait}
Johnson, R.~C. and Noguera, G. (2017).
\newblock A portrait of trade in value-added over four decades.
\newblock {\em Review of Economics and Statistics}, 99(5):896--911.

\bibitem[Kasahara and Rodrigue, 2008]{kasahara2008does}
Kasahara, H. and Rodrigue, J. (2008).
\newblock Does the use of imported intermediates increase productivity?
  plant-level evidence.
\newblock {\em Journal of development economics}, 87(1):106--118.

\bibitem[Keller, 2002]{keller2002trade}
Keller, W. (2002).
\newblock Trade and the transmission of technology.
\newblock {\em Journal of Economic growth}, 7(1):5--24.

\bibitem[Kugler and Verhoogen, 2009]{kugler2009plants}
Kugler, M. and Verhoogen, E. (2009).
\newblock Plants and imported inputs: New facts and an interpretation.
\newblock {\em American Economic Review}, 99(2):501--07.

\bibitem[Kugler and Verhoogen, 2011]{kugler2011prices}
Kugler, M. and Verhoogen, E. (2011).
\newblock Prices, plant size, and product quality.
\newblock {\em The Review of Economic Studies}, 79(1):307--339.

\bibitem[Lin, 2010]{lin2010identifying}
Lin, X. (2010).
\newblock Identifying peer effects in student academic achievement by spatial
  autoregressive models with group unobservables.
\newblock {\em Journal of Labor Economics}, 28(4):825--860.

\bibitem[L{\'o}pez and Yadav, 2010]{lopez2010imports}
L{\'o}pez, R.~A. and Yadav, N. (2010).
\newblock Imports of intermediate inputs and spillover effects: Evidence from
  chilean plants.
\newblock {\em The Journal of Development Studies}, 46(8):1385--1403.

\bibitem[Lu et~al., 2024]{lu2024firms}
Lu, D., Mariscal, A., and Mej{\'\i}a, L.-F. (2024).
\newblock How firms accumulate inputs: Evidence from import switching.
\newblock {\em Journal of International Economics}, 148:103847.

\bibitem[Manski, 1993]{manski1993identification}
Manski, C.~F. (1993).
\newblock Identification of endogenous social effects: The reflection problem.
\newblock {\em The review of economic studies}, 60(3):531--542.

\bibitem[Melitz, 2003]{melitz2003impact}
Melitz, M.~J. (2003).
\newblock The impact of trade on intra-industry reallocations and aggregate
  industry productivity.
\newblock {\em Econometrica}, 71(6):1695--1725.

\bibitem[Rauch and Watson, 2003]{Rauch2003}
Rauch, J. and Watson, J. (2003).
\newblock Starting small in an unfamiliar environment.
\newblock {\em International Journal of Industrial Organization},
  21:1021--1042.

\bibitem[Rauch, 1999]{rauch1999networks}
Rauch, J.~E. (1999).
\newblock Networks versus markets in international trade.
\newblock {\em Journal of international Economics}, 48(1):7--35.

\bibitem[Rauch and Watson, 2004]{rauch2004network}
Rauch, J.~E. and Watson, J. (2004).
\newblock Network intermediaries in international trade.
\newblock {\em Journal of Economics \& Management Strategy}, 13(1):69--93.

\end{thebibliography}

\newpage
\section*{Appendix}

\subsection*{Tables}
\vspace{20pt}

% \begin{center}
%     ******** NEW TABLE RESTRICTIVE STRATEGY OLS ********
% \end{center}
\begin{table}[H]
	\begin{center}
	\caption{OLS results}
  \vspace{5pt}
	\label{tab: olss1}
	%{\footnotesize
		%\rowcolors{1}{}{lightgray}
  \scalebox{0.8}{
		\begin{tabular}{lccHccccH} \hline\hline
           &   {\color{black}(1)}   &    {\color{black}(2)}   &    {\color{black}(3)}   &    {\color{black}(3)}   &    {\color{black}(4)}   &    {\color{black}(5)} &    {\color{black}(6)}   \\  \hline
$\bar{y}^{D}_{ihc, t-1}$ &  0.0314***   &  0.0316***    &   0.0778***   &  0.0302***    &    0.0142***  &  0.0118***    &    0.0003  \\
   &  (0.0007)     &   (0.0007)    &    (0.0025)  &  (0.0007)     &  (0.0007)     &   (0.0009)   &  (0.0023)     \\
$\bar{y}^{U}_{ihc, t-1}$ &  0.0317***    & 0.0319***     &  0.0671***  &   0.0303***    &    0.0125***   &      0.0102***&   0.00125   \\
            &  (0.0008)     &   (0.0007)    &  (0.0025)   &   (0.0007)     &   (0.0007)    &    (0.0011)   &  (0.0019)    \\
$\text{\textit{prop\_imp\_zip}}_{ihc,t-1}$    &               &               &               &               &      0.1886***&               &               \\
            &               &               &               &               &    (0.0037)   &               &               \\
$\text{\textit{prop\_imp\_sec}}_{ihc,t-1}$    &               &               &               &               &      0.1966***&               &               \\
            &               &               &               &               &    (0.0042)   &               &               \\
$\text{\textit{prop\_imp\_sec\_zip}}_{ihc,t-1}$  &               &               &               &               &      0.0363***&               &               \\
            &               &               &               &               &    (0.0024)   &               &               \\
$\text{Own/Peers' characteristics}$        &      No         &   Yes            &     Yes   &      No         &     No          &    No           &  Yes  \\ 
\hline
r2          &  0.2719       &   0.2726      &    0.3039     &   0.5399      &     0.5458    &      0.6574   &  0.5721       \\
N           & 2048865    &  2048865    &  1105092    &  1702966    &    1702966   &  1238540    &  537280    \\
fixed effects     & \textit{id}   & \textit{id}   & \textit{id}    &\textit{id-y}    &\textit{id-y}    &\textit{id-y}    &\textit{id}    \\
               &  \textit{eu-y}   &  \textit{eu-y}   &  \textit{eu-y}   & \textit{eu-y}   & \textit{eu-y}   & \textit{eu-s-z-y}   & \textit{s-z-y}   \\
clustering variable    &         \textit{id}   &         \textit{id}   &         \textit{id}   &    \textit{id-y}   &    \textit{id-y}   &   \textit{id-y}   &    \textit{id}   \\
	\hline\hline
		\end{tabular}
  }
%}
\end{center}
\vspace{-5pt}
\begin{spacing}{1}
{\footnotesize\textit{Notes}:  The dependent variable is a dummy equal to one if a firm $i$ belonging to \textit{industry$\times$zip\_code} $h$ starts importing from country $c$ at year $t$;  $id$ refers to the firm identification code; \textit{eu-y} refers to \textit{import\_origin$\times$year} fixed effects; \textit{id-y} refers to \textit{firm$\times$year} fixed effects; \textit{eu-s-z-y} refers to \textit{import\_origin $\times$sector$\times$zip\_code$\times$year} fixed effects. $^{*}$p$<$0.1;$^{**}$p$<$0.05; $^{***}$p$<$0.01. }
\end{spacing}

\end{table}
\newpage
%%%%%%%%%%%%%%%%%%% IV REGRESSION:

\begin{table}[H]
	\begin{center}
	\caption{IVs results}
 \vspace{5pt}
	\label{tab: IV_fixed_restrictive}
	
		%\rowcolors{1}{}{lightgray}
        \scalebox{0.8}{
		\begin{tabular}{lcccc} \hline\hline
           &   {(1)}   &    {(2)}   &    {(3)} &    {(4)}\\  \hline 
$\bar{y}^{D}_{ihc, t-1}$ &      0.0386***&    0.0561***  & 0.4314* & 0.4334\\
            &    (0.0141)   &   (0.0180)    & (0.2133) & (0.2989)\\
$\bar{y}^{U}_{ihc, t-1}$ &  0.0684***  &   0.0881*** & -0.0899 & 0.0984\\
            &  (0.0205)    &   (0.0234)  & (0.1627) & (0.1334)\\
Own/Peers' characteristics &  No   &  No  & Yes & Yes \\
\hline
r2          &      0.6568   &      0.6012   & -0.0982 & 0.3369\\
N           & 780210   & 501566  & 531893 & 338016\\
idstat      &   932.396   &   659.486  & 20.303 & 10.642\\
idp         &      0.0000   &      0.0000  & 0.0000 & 0.0138\\
widstat     &   361.510   &    127.806  & 27.392 & 11.04\\
j           &         &      1.338  &   & 2.915\\
jp          &               &     0.5121   &  & 0.2328\\
instruments & $\bar{\bar{y}}^{D}_{ihc, t-2}$  & $\bar{\bar{y}}^{D}_{ihc, t-2}$   & $\bar{\bar{y}}^{D}_{ihc, t-2}$  & $\bar{\bar{y}}^{D}_{ihc, t-2}$ \\
& $\bar{\bar{y}}^{U}_{ihc, t-2}$  & $\bar{\bar{y}}^{U}_{ihc, t-2}$ &  $\bar{\bar{y}}^{U}_{ihc, t-2}$ & $\bar{\bar{y}}^{U}_{ihc, t-2}$\\
%                 &  & &    \\
                 &                  & $\bar{\bar{y}}^{D}_{ihc, t-3}$ & & $\bar{\bar{y}}^{D}_{ihc, t-3}$    \\
                 &                  & $\bar{\bar{y}}^{U}_{ihc, t-3}$  &  & $\bar{\bar{y}}^{U}_{ihc, t-3}$  \\
                 
fixed effects     &id-y     &id-y   & id  & id    \\
               &eu-s-z-y    &eu-s-z-y   &  s-z-y &  s-z-y  \\
clustering variable    &    id-y  &    id-y   &  id & id \\
\hline\hline
		\end{tabular}
}
\end{center}
\vspace{-5pt}
\begin{spacing}{1}
		{\footnotesize \textit{Notes}: The dependent variable in columns 1 and 2 is a dummy equal to one if a firm $i$ belonging to \textit{industry$\times$zip\_code} $h$ starts importing from country $c$ at year $t$. The dependent variable in columns 3 and 4 is a dummy equal to one if a firm $i$ belonging to \textit{industry$\times$zip\_code} $h$ starts importing at year $t$. \textit{id} refers to \textit{firm} fixed effects; \textit{id-y} refers to \textit{firm$\times$year} fixed effects; \textit{eu-s-z-y} refers to \textit{import\_origin$\times$industry$\times$zip\_code$\times$year} fixed effects; idstat refers to the underidentification test (Kleibergen-Paap rk LM statistic; under the null the equation is underidentified); idp is the p-value corresponding to idstat; widstat refers to the weak identification test (Kleibergen-Paap rk Wald F statistic; under the null the IVs are weak, Stock and Yogo (2005)); j refers to the overidentification test of all instruments (Hansen J statistic; under the null the IVs are uncorrelated with the error); jp is the p-value of j. $^{*}$p$<$0.1;$^{**}$p$<$0.05; $^{***}$p$<$0.01.}
	
 \end{spacing}
\end{table}
\newpage
\begin{table}[H]
\begin{center}
\caption{Heterogeneity of peer effect by firm characteristics}\label{tbl:het_firm_char}
 \vspace{5pt}
\scalebox{0.8}{
\begin{tabular}{lccccHc}
\hline
\hline
  & \makecell{Number of \\ Workers} & \makecell{Number of \\  Suppliers} & \makecell{Number of \\ Customers} & \makecell{Labor \\ Productivity} & \makecell{Intermediate Inputs \\ Productivity} & \makecell{Being \\ a Wholesaler}\\
%& & (1) & (2) & (3) \\
\hline
$\bar{y}^{D}_{ic, t-1}*z_{i}^{Low}$ & 0.0045***  & 0.0018 & 0.0010  & 0.0039***  & 0.0060*** &0.0088***\\
 & (0.0013) & (0.0011)  & (0.0013)& (0.0014)&  (0.0016)& (0.0012)\\
 $\bar{y}^{D}_{ic, t-1}*z_{i}^{High}$  & 0.0213*** & 0.0479***  & 0.0189*** & 0.0173*** & 0.0133***&0.0179***\\
 &  (0.0018) & (0.0025) & (0.0016) & (0.0016) & (0.0014)& (0.0034)\\
\hline
$\bar{y}^{U}_{ic, t-1}*z_{i}^{Low}$& 0.0061***  & 0.0030***  & 0.0023** & 0.0078***& 0.0071***& 0.0097***\\
 & (0.0012)  & (0.0011) & (0.0011) & (0.0013) & (0.0014)& (0.0010)\\
 $\bar{y}^{U}_{ic, t-1}*z_{i}^{High}$ & 0.0200*** & 0.0259***  & 0.0281*** & 0.0162*** & 0.0157***& 0.0227***\\
& (0.0016)   & (0.0018)  & (0.0017)  & (0.0014) & (0.0013)&(0.0030)\\
\hline
%* Group indicators & & Yes & Yes & Yes & Yes&Yes\\
%Firm-year FE & & Yes & Yes & Yes & Yes&Yes\\
%Country-year FE & & Yes & Yes & Yes & Yes&Yes\\
$N$  & 1,238,540 & 1,238,540 & 1,238,540 & 1,238,540 & 1,238,540 & 1,238,540\\
\multirow{2}{*}{fixed effects}   & id-y & id-y & id-y & id-y& id-y& id-y\\
 & eu-s-z-y & eu-s-z-y & eu-s-z-y& eu-s-z-y&eu-s-z-y &eu-s-z-y\\
clustering variable &id-y  & id-y &  id-y& id-y& id-y & id-y \\
	\hline
\hline
\end{tabular} }
\end{center}
\vspace{-7pt}
\begin{spacing}{1}
{\footnotesize Note: The dependent variable is a dummy equal to one if a firm $i$ belonging to \textit{industry$\times$zip code} $h$ starts importing from country $c$ at year $t$. $\bar{y}^{D}_{ic, t-1}$ and $\bar{y}^{U}_{ic, t-1}$ denote the number of importing suppliers or customers of firm $i$ divided by the total number of suppliers or customers of firm $i$, respectively. $z_{i}^{Low}$ ($z_{i}^{High}$) refer to an indicator variable for having the value of the characteristic at the top of the column below (above) the observed median value of that characteristic. In the last column, High (Low) means that the firm is (not) a wholesaler. \textit{id-y} refers to firm$\times$year fixed effects; \textit{eu-s-z-y} refers to import origin$\times$industry$\times$zip code$\times$year fixed effects. $^{*}$p$<$0.1;$^{**}$p$<$0.05; $^{***}$p$<$0.01.}
\end{spacing}
\end{table}

\newpage
\begin{table}[H]
\begin{center}
\caption{Heterogeneity of peer effect by peers characteristics}\label{tbl:het_peer_char}
 \vspace{5pt}
\scalebox{0.8}{
\begin{tabular}{lccccHc}
\hline
\hline
  & \makecell{Number of \\ Workers} & \makecell{Number of \\  Suppliers} & \makecell{Number of \\  Customers} & \makecell{Labor \\ Productivity} & \makecell{Intermediate Inputs \\ Productivity}& \makecell{Wholesalers}\\
%& & (1) & (2) & (3) \\
\hline
\multirow{2}{*}{$\bar{y}^{D, Low}_{ic,t-1}$}& 0.0173***  & 0.0308*** &  0.0406*** & 0.0185***& 0.0091*** & 0.0187***\\
 &(0.0027)& (0.0045) & (0.0120) & (0.0025) & (0.0026) &  (0.0016)\\
\multirow{2}{*}{$\bar{y}^{D, High}_{ic,t-1}$}& 0.0090***& 0.0087*** & 0.0099*** &  0.0084***& 0.0105*** & 0.0021\\
& (0.0012) &(0.0011)  & (0.0011) &  (0.0012)&(0.0012)  & (0.0015) \\
\hline
\multirow{2}{*}{$\bar{y}^{U, Low}_{ic,t-1}$}& 0.0190*** & 0.0344*** & 0.0181*** & 0.0143***& 0.0167*** & 0.0123***\\
 &(0.0026)& (0.0055) & (0.0025)  & (0.0019) & (0.0023)  & (0.0012) \\
\multirow{2}{*}{$\bar{y}^{U, High}_{ic,t-1}$}& 0.0106*** & 0.0109***& 0.0105*** & 0.0110*** & 0.0108***& 0.0101*** \\
& (0.0011) & (0.0010) & (0.0011) & (0.0011) & (0.0011) & (0.0017) \\
\hline
%* Group indicators & & Yes & Yes & Yes & Yes&Yes\\
%Firm-year FE & & Yes & Yes & Yes & Yes&Yes\\
%Country-year FE & & Yes & Yes & Yes & Yes&Yes\\
$N$ & 1,238,540&  1,238,540& 1,238,540 & 1,238,540 &1,238,540 &1,238,540 \\
\multirow{2}{*}{fixed effects}  &id-y& id-y & id-y & id-y &id-y &id-y \\
 &eu-s-z-y& eu-s-z-y & eu-s-z-y &eu-s-z-y &eu-s-z-y&eu-s-z-y  \\
clustering variable &id-y&  id-y& id-y &  id-y&  id-y & id-y  \\
	\hline
\hline
\end{tabular} 
}
\end{center}
\vspace{-7pt}
\begin{spacing}{1}
{\footnotesize \textit{Notes}: The dependent variable is a dummy equal to one if a firm $i$ belonging to \textit{industry$\times$zip code} $h$ starts importing from country $c$ at year $t$. $\bar{y}^{D,v}_{ihc,t-1}$ and  $\bar{y}^{U,v}_{ihc,t-1}$  for a given firm $i$ denote the number of importing suppliers or customers of firm $i$ in category $v$ (i.e., with the value of the considered characteristic lower or higher than the median, Low and High respectively) divided by the total number of suppliers or customers of firm  $i$, respectively. In the last column, High (Low) means that the numerator of the treatment variables counts only the neighbours that are (not) wholesalers. \textit{id-y} refers to firm$\times$year fixed effects; \textit{eu-s-z-y} refers to import origin$\times$industry$\times$zip code$\times$year fixed effects. $^{*}$p$<$0.1;$^{**}$p$<$0.05; $^{***}$p$<$0.01.}
\end{spacing}

\end{table}

\newpage
\begin{table}[H]
\begin{center}
\caption{Heterogeneity of peer effect by firm characteristics and peers characteristics (1a)}\label{tbl:het_firmpeer_char}
 \vspace{5pt}
\scalebox{0.8}{
\begin{tabular}{lccccHc}
\hline
\hline
 &  \makecell{Number of \\ Workers} & \makecell{Number of \\ Suppliers} & \makecell{Number of \\  Customers} & \makecell{Labor \\ Productivity} & \makecell{Intermediate Inputs \\ Productivity} & \makecell{Wholesalers}\\
%& & (1) & (2) & (3) \\
\hline
$\bar{y}^{D}_{ihc,t-1}$ & 0.0091***  &  0.0168***  & 0.0157   &  0.0202***   &  0.0064* & 0.0182*** \\
 & (.0030) & (0.0046)  & (0.0153)& (0.0037)&  (0.0034)& (0.0016)\\
 
$\bar{y}^{D}_{ihc,t-1}*z_{i}^{High}$    & 0.0299*** & 0.1512***   &  0.0492** &  -0.0034  & 0.0056 & 0.0042\\
& (0.0064) & (0.0188) & (0.0237) & (0.0049) & (0.0047)& (0.0054)\\

$\bar{y}^{D, High}_{ihc,t-1}$ & -0.0056*  &  -0.0163*** &  -0.0150   & -0.0195***   &   -0.0007  & -0.0191*** \\
 & (0.0033) & (0.0047)  & (0.0153)& (0.0039)&  (0.0038)& (0.0022)\\
 
 $\bar{y}^{D, High}_{ihc,t-1}*z_{i}^{High}$ & -0.0146** &  -0.1083*** & -0.0317 & 0.0200*** & 0.0022 &  0.0114*\\
&  (0.0068) & (0.0191) & (0.0237) & (0.0053) & (0.0052)& (0.0068)\\

\hline

$\bar{y}^{U}_{ihc,t-1}$& 0.0145***  & 0.0252***   & 0.0091*** & 0.0117*** & 0.0158*** & 0.0107*** \\
 & (0.0029)  & (0.0061) & (0.0030) & (0.0025) & (0.0030)& (0.0012)\\
 
$\bar{y}^{U}_{ihc,t-1}*z_{i}^{High}$ & 0.0147** & 0.0331**   &  0.0243*** & 0.0060  & 0.0023 & 0.0150*** \\
 & (0.0060)   & (0.0134)  & (0.0052)  & (0.0038) & (0.0046)&(0.0044)\\

$\bar{y}^{U, High}_{ihc,t-1}$ & -0.0101***  & -0.0232***   & -0.0081**  & -0.0054* &  -0.0111***& -0.0049** \\
& (0.0031)  & (0.0062) & (0.0032) & (0.0029) & (0.0033)& (0.0022)\\

 $\bar{y}^{U, High}_{ihc,t-1}*z_{i}^{High}$ & -0.0003 &   -0.0101 & 0.0016 & 0.0034 & 0.0085* & -0.0005\\
&  (0.0063)   & (0.0136)  & (0.0056)  & (0.0044) & (0.0051) & (0.0060)\\
\hline
%* Group indicators & & Yes & Yes & Yes & Yes&Yes\\
%Firm-year FE & & Yes & Yes & Yes & Yes&Yes\\
%Country-year FE & & Yes & Yes & Yes & Yes&Yes\\
$N$  & 1,238,540 & 1,238,540 & 1,238,540 & 1,238,540 & 1,238,540 & 1,238,540\\
\multirow{2}{*}{fixed effects}    & id-y & id-y & id-y & id-y& id-y& id-y\\
 & eu-s-z-y & eu-s-z-y & eu-s-z-y& eu-s-z-y&eu-s-z-y &eu-s-z-y\\
clustering variable &id-y  & id-y &  id-y& id-y& id-y & id-y \\
	\hline
\hline
\end{tabular} }
\end{center}
\vspace{-7pt}
\begin{spacing}{1}
\footnotesize {\textit{Notes}: The dependent variable is a dummy equal to one if a firm $i$ belonging to \textit{industry$\times$zip code} $h$ starts importing from country $c$ at year $t$. $\bar{y}^{D}_{ic, t-1}$ and $\bar{y}^{U}_{ic, t-1}$ denote the number of importing suppliers or customers of firm $i$ divided by the total number of suppliers or customers of firm $i$, respectively. $\bar{y}^{D, High}_{ihc,t-1}$ and  $\bar{y}^{U, High}_{ihc,t-1}$ denote the number of importing suppliers or customers of firm $i$ having the value of the considered characteristic higher than the observed median value of that characteristic divided by the total number of suppliers or customers of firm  $i$, respectively. $z_{i}^{High}$ is an indicator variable for having the value of the considered characteristic above the observed median value of that characteristic. In the last column, the numerator of the variables $\bar{y}^{D, High}_{ihc,t-1}$ and  $\bar{y}^{U, High}_{ihc,t-1}$ counts only the neighbours that are wholesalers and $z_{i}^{High}$ is an indicator variable for firm $i$ being a wholesaler. \textit{id-y} refers to firm$\times$year fixed effects; \textit{eu-s-z-y} refers to import origin$\times$industry$\times$zip code$\times$year fixed effects. $^{*}$p$<$0.1;$^{**}$p$<$0.05; $^{***}$p$<$0.01.}
\end{spacing}

\end{table}

\newpage
\begin{table}[H]
\begin{center}
\caption{Heterogeneity of peer effect by firm characteristics and peers characteristics (1b)}\label{tbl:het_firmpeer_char_groups}
 \vspace{5pt}
\scalebox{0.7}{
\begin{tabular}{lcccccc}
\hline
\hline
 & & \makecell{Number of \\ Workers} & \makecell{Number of \\ Suppliers} & \makecell{Number of \\  Customers} & \makecell{Labor \\ Productivity}  & \makecell{Wholesalers}\\
%& & (1) & (2) & (3) \\
\hline
\multirow{4}{*}{$\bar{y}^{D, Low}_{ihc,t-1}$} &Low& 0.0091***  & 0.0168***  & 0.0157   & 0.0202***   &   0.0182*** \\
 && (0.0030) & (0.0046)  & (0.0153)& (0.0037)&   (0.0016)\\
 &High & 0.0390*** & 0.1680***   &  0.0649*** & 0.0167***   & 0.0225***\\
& &  (0.0057) & (0.0183) & (0.0181) & (0.0033) &  (0.0052)\\
\hline
\multirow{4}{*}{$\bar{y}^{D, High}_{ihc,t-1}$} &Low& 0.0035**  &  0.0005 & 0.0008   & 0.0006     &-0.0009 \\
 && (0.0014) & (0.0012)  & (0.0013)& (0.0015)&   (0.0015)\\
 &High & 0.0188*** &  0.0434*** & 0.0183*** & 0.0172*** &   0.0148***\\
& &  (0.0019) & (0.0025) & (0.0016) & (0.0017) & (0.0042)\\
\hline
\multirow{4}{*}{$\bar{y}^{U, Low}_{ihc,t-1}$} &Low& 0.0145***  & 0.0252***   & 0.0091*** & 0.0117***  & 0.0107*** \\
 && (0.0029)  & (0.0061) & (0.0030) & (0.0025) & (0.0012)\\
 &High & 0.0292*** & 0.0583***   &  0.0334*** & 0.0177***   & 0.0257*** \\
& & (0.0052)   & (0.0120)  & (0.0043)  & (0.0029) &(0.0042)\\
\hline
\multirow{4}{*}{$\bar{y}^{U, High}_{ihc,t-1}$} &Low& 0.0045***  & 0.0019*   & 0.0010  & 0.0062*** &   0.0058*** \\
 && (0.0013)  & (0.0011) & (0.0012) & (0.0015) &  (0.0018)\\
 &High & 0.0188*** &  0.0248*** & 0.0271*** & 0.0156*** & 0.0203***\\
& & (0.0017)   & (0.0018)  & (0.0019)  & (0.0016) &(0.0040)\\
\hline
%* Group indicators & & Yes & Yes & Yes & Yes&Yes\\
%Firm-year FE & & Yes & Yes & Yes & Yes&Yes\\
%Country-year FE & & Yes & Yes & Yes & Yes&Yes\\
$N$ &  & 1,238,540 & 1,238,540 & 1,238,540 & 1,238,540 & 1,238,540 \\
\multirow{2}{*}{fixed effects}  &  & id-y & id-y & id-y & id-y& id-y\\
 && eu-s-z-y & eu-s-z-y & eu-s-z-y& eu-s-z-y&eu-s-z-y \\
clustering variable &&id-y  & id-y &  id-y& id-y& id-y  \\
	\hline
\hline
\end{tabular} }
\end{center}
\vspace{-7pt}
\begin{spacing}{1}
\footnotesize {\textit{Notes}: This table presents the results, which are obtained when allowing interaction effects between firm and peer characteristics (and are therefore equivalent to those reported in Table \ref{tbl:het_firmpeer_char}), by considering the four types of interactions: low/high characteristics of firms with low/high characteristics of peers. In this way, instead of presenting the results as incremental effects with respect to a baseline category as in Table \ref{tbl:het_firmpeer_char} (i.e., the coefficient associated to $\beta_{S}^{L,L}$ in \ref{eq:regression_het_firm_peer} measures the spillover effect to a small firm from small importing peers; notice that the first and fifth rows of this Table and those of Table \ref{tbl:het_firmpeer_char} are the same), we show the estimated effects of peer spillovers based on whether both a firm's characteristic and its peers’ characteristic lie above or below the median, or whether they are in opposite parts of the distribution. Therefore, the estimates presented in this Table are obtained by interacting the $z_{i,t}^{v}$ variables with the $\bar{y}^{S, v}_{ihc,t-1}$ variables (with $v \in \{L,H\}$ and $S \in \{D,U\}$). $\bar{y}^{D, v}_{ihc,t-1}$ and  $\bar{y}^{U, v}_{ihc,t-1}$  for a given firm $i$ denote the number of importing suppliers and customers of firm $i$ in category $v$ (i.e., with the value of the considered characteristic lower or higher than the median, Low and High respectively) divided by the total number of suppliers and customers of firm  $i$, respectively. In the last column, High (Low) means that the numerator of the treatment variables counts only the neighbours that are (not) wholesalers. Low (high) in the second column refers to the interaction of the variables $\bar{y}^{D, v}_{ihc,t-1}$ and  $\bar{y}^{U, v}_{ihc,t-1}$ with an indicator variable $z_{i,t}^{Low}$ ($z_{i,t}^{High}$) for firm $i$ having the value of the characteristic at the top of the column below (above) the observed median value of that characteristic. In the last column, High (Low) means that the firm is (not) a wholesaler. \textit{id-y} refers to firm$\times$year fixed effects; \textit{eu-s-z-y} refers to import origin$\times$sector$\times$zip code$\times$year fixed effects. $^{*}$p$<$0.1;$^{**}$p$<$0.05; $^{***}$p$<$0.01.}
\end{spacing}

\end{table}

\begin{table}[H]
\begin{center}
\caption{Heterogeneity of peer effect by firm characteristics and peers characteristics (2)}\label{tbl:het_link}
 \vspace{5pt}

\scalebox{0.8}{
\begin{tabular}{lcccc}
\hline
\hline
  & \makecell{Same \\ Industry} & \makecell{Same \\ ZIP code} & \makecell{Same \\ Province} & \makecell{Reciprocal \\ Relationship}\\
%& & (1) & (2) & (3) \\
\hline
\multirow{2}{*}{$\bar{y}^{D, No}_{ihc,t-1}$}& 0.0092***   & 0 .0115***  & 0.0195***   &0.0103***  \\
 & (0.0012) &  (0.0012) & 0.0018)  & (0.0011)  \\
\multirow{2}{*}{$\bar{y}^{D, Yes}_{ihc,t-1}$}& 0.0177*** & 0.0044  &  0.0048***  &  0.0217***  \\
& (0.0034) & (0.0026)  & (0.0014)  & (0.0037)  \\
\hline
\multirow{2}{*}{$\bar{y}^{U, No}_{ihc,t-1}$}& 0.0109***  &  0.0138*** & 0.0194***  & 0.0107***  \\
 & (0.0011) & (0.0011)  &  (0.0018)  &  (0.0010) \\
\multirow{2}{*}{$\bar{y}^{U, Yes}_{ihc,t-1}$}&  0.0161*** & 0.0041*  &  0.0083*** &  collinear \\
&  (0.0027) &  (0.0021) &  (0.0012) &  with $S_{ic, t-1}^{Yes}$  \\
\hline
%* Group indicators & & Yes & Yes & Yes & Yes&Yes\\
%Firm-year FE & & Yes & Yes & Yes & Yes&Yes\\
%Country-year FE & & Yes & Yes & Yes & Yes&Yes\\
$N$ & 1,238,540&  1,238,540& 1,238,540 & 1,238,540  \\
\multirow{2}{*}{fixed effects}  &id-y& id-y & id-y & id-y  \\
 &eu-s-z-y& eu-s-z-y & eu-s-z-y &eu-s-z-y   \\
clustering variable &id-y&  id-y& id-y &  id-y  \\
	\hline
\hline
\end{tabular} 
}
\end{center}
\vspace{-7pt}
\begin{spacing}{1}
\footnotesize {\textit{Notes}:The dependent variable is a dummy equal to one if a firm $i$ belonging to \textit{industry$\times$zip code} $h$ starts importing from country $c$ at year $t$. $\bar{y}^{D,v}_{ihc,t-1}$ and  $\bar{y}^{U,v}_{ihc,t-1}$  for a given firm $i$ denote the number of importing suppliers and customers of firm $i$ in category $v$ divided by the total number of suppliers and customers of firm  $i$, respectively. $v=Yes$ ($v=No$) refers to the group of peers that (do not) belong to the same industry or are (not) located in the same zip code or province or (do not) form a reciprocal relationship with firm $i$. \textit{id-y} refers to firm$\times$year fixed effects; \textit{eu-s-z-y} refers to import origin$\times$industry$\times$zip code$\times$year fixed effects. $^{*}$p$<$0.1;$^{**}$p$<$0.05; $^{***}$p$<$0.01.}
\end{spacing}
\end{table}

\newpage

\begin{table}[htp]
\vspace{-10mm}
\begin{center}
\caption{Descriptive statistics: characteristics of firms}
\label{tbl:destab1}
 \vspace{5pt}
{\small
%\rowcolors{1}{}{lightgray}
\begin{tabular}{lccc}%{l @{\hspace*{10mm}} c @{\hspace*{10mm}} c @{\hspace*{10mm}} c} 
\hline\hline
	& (1)				& (2)				& (3)	\\
	& EU-Starters  	& non EU-Starters		&  non-Starters\\
&  premium  	&  premium		&  baseline\\
    
\hline
\symbol{35} Workers     &       6.3***&       10.8***&       9.6***\\
            &     (0.5)&     (1.1)&     (0.1)\\
\symbol{35} Dom. suppliers     &            5.4***&        3.7***&       5.7***\\
            &         (0.1)&     (0.2)&     (0.0)\\
\symbol{35} Dom. customers &   4.6***    &  3.4***     & 5.5***      \\
            &     (0.2)&     (0.2)&     (0.0)\\
Int. input cost &    474.2***    &  592.3***      &   309.9***    \\
            &     (32.3)&     (66.9)&     (6.3)\\
Total sales &    1271.4***    &     1442.2***   &     921.7***  \\
            &     (87.1)&     (141.5)&     (10.7)\\
Sales to firms&     474.1***    &  643.8***    &  289.0***    \\
            &     (43.1)&     (114.1)&     (6.4)\\
Domestic sales&    1235.2***    &  1407.1***    &    918.4***    \\
            &     (86.8)&     (140.7)&     (10.7)\\
Sales per customer&    51.1***    &   31.3***     &    95.4***   \\
            &     (9.7)&     (7.1)&     (1.3)\\
Labor productivity&  66.8***    &  31.1***    &  178.2***    \\
            &     (10.2)&     (9.8)&     (1.4)\\
%Int. input productivity&  22.2    &  19.8**   &  191.7***  \\
 %           &    (13.9)&    (8.4)&    (2.0)\\
Avg. labor cost&   1.5***  &  1.6  &  28.1***  \\
            & (0.5)& (1.0)& (0.1)\\

\hline
Number of firms	&       30320&       13497&       142705\\

\hline\hline
\end{tabular}}
\end{center}
\vspace{-5pt}
\begin{spacing}{1}
{\footnotesize \textit{Notes}: Descriptive statistics for the year 2010 on the sample used for specification (5) of Table \ref{tab: olss1}. Monetary variables are in thousands of euros. We report the estimated coefficients obtained by regressing one by one the relevant characteristics on a constant, a dummy for being an import starter from some EU country and a dummy for being an import starter from some non-EU country. Int. input cost is the total cost of inputs from domestic providers. Sales to firms are the value of sales to other domestic firms.  Domestic sales are sales to domestic firms and non-firms. Sales per customer are the average domestic sales per domestic customer. Labor productivity is the ratio between total sales and the number of workers. Int. input productivity is the ratio between total sales and the number of domestic suppliers. Avg. labor cost is total labor cost divided by the number of workers. $^{*}$p$<$0.1;$^{**}$p$<$0.05; $^{***}$p$<$0.01.}
\end{spacing}
\end{table}

\newpage

\begin{table}[htp]
\vspace{-10mm}
\begin{center}
\caption{Descriptive statistics: import behavior of peers }
\label{tbl:destab2}
 \vspace{5pt}
{\small
%\rowcolors{1}{}{lightgray}
\begin{tabular}{lccc}%{l @{\hspace*{10mm}} c @{\hspace*{10mm}} c @{\hspace*{10mm}} c} 
\hline\hline
	& (1)				& (2)				& (3)	\\
	& EU-Starters  	& non EU-Starters		&  non-Starters\\
&  premium  	&  premium		&  baseline\\
    
\hline
\symbol{35} Dom. suppliers     &            3.9***&        2.0***&       3.2***\\
importing from EU            &         (0.1)&     (0.1)&     (0.0)\\
\symbol{35} Dom. customers &   2.6***    &  2.1***     & 2.3***      \\
importing from EU            &     (0.1)&     (0.1)&     (0.0)\\
\symbol{35} Dom. suppliers     &            2.6***&        2.0***&       2.2***\\
importing from non EU            &         (0.1)&     (0.1)&     (0.0)\\
\symbol{35} Dom. customers &   1.5***    &  2.1***     & 1.4***      \\
importing from non EU            &     (0.1)&     (0.1)&     (0.0)\\

Proportion Dom. suppliers     &      0.09***&        0.01***&       0.51***\\
importing from EU            &         (0.00)&     (0.00)&     (0.00)\\
Proportion Dom. customers &   0.09***    &  0.04***     & 0.46***      \\
importing from EU            &     (0.00)&     (0.00)&     (0.00)\\
Proportion Dom. suppliers     &       0.05***&        0.11***&       0.35***\\
importing from non EU            &         (0.00)&     (0.00)&     (0.00)\\
Proportion Dom. customers &   0.02***    &  0.13***     & 0.33***      \\
importing from non EU            &     (0.00)&     (0.00)&     (0.00)\\
\hline
Number of firms	&       30320&       13497&       142705\\

\hline\hline
\end{tabular}}
\end{center}
\vspace{-5pt}
\begin{spacing}{1}
{\footnotesize \textit{Notes}: Descriptive statistics for the year 2010 on the sample used for specification (5) of Table \ref{tab: olss1}. Monetary variables are in thousands of euros. We report the estimated coefficients obtained by regressing one by one the relevant characteristics on a constant, a dummy for being an import starter from some EU country and a dummy for being an import starter from some non-EU country. $^{*}$p$<$0.1;$^{**}$p$<$0.05; $^{***}$p$<$0.01.}
\end{spacing}
\end{table}

\newpage
\begin{table}[htbp]
\begin{center}
    \caption{Observed probability to start importing by number of workers quintile}\label{tbl:quintile_size_import}
     \vspace{5pt}
    \begin{tabular}{cc}
        \hline \hline
%\vspace{-10pt}
        Probability& Quintile \\
        \midrule
        0.14 & 1 \\
        0.19 & 2 \\
        0.21 & 3 \\
        0.24 & 4 \\
        0.31 & 5 \\
        \hline \hline
    \end{tabular}
    \end{center}
    \vspace{-7pt}
    \begin{spacing}{1}
    {\footnotesize \textit{Notes}: Row 1 reports the share of firms that start to import having the number of workers lower than the first quintile of the distribution of number of workers in the sample used to estimate specification (5) of Table \ref{tab: olss1}. Other entries have analogous interpretations.  Starters are firms that start to import after 2010.}
    \end{spacing}
\end{table}

\end{document}